\documentclass[longauth,structabstract]{aa}

\usepackage{txfonts}
\usepackage{graphicx}
\usepackage{natbib}
\usepackage{multirow}
\usepackage{rotating}
\usepackage{hyperref}

\bibpunct{(}{)}{;}{a}{}{,}

\begin{document}

\title{The Simultaneous Low State Spectral Energy Distribution of 1ES\,2344$+$514 from Radio to Very High Energies}
%
\author{
 J.~Aleksi\'c\inst{1} \and
 L.~A.~Antonelli\inst{2} \and
 P.~Antoranz\inst{3} \and
 M.~Asensio\inst{4} \and
 M.~Backes\inst{5} \and
 U.~Barres de Almeida\inst{6} \and
 J.~A.~Barrio\inst{4} \and
 W.~Bednarek\inst{7} \and
 K.~Berger\inst{8,}\inst{9} \and
 E.~Bernardini\inst{10}
% ...
\\}

\offprints{S.~ R\"ugamer, \email{snruegam@astro.uni-wuerzburg.de}, E.~Lindfors, \email{elilin@utu.fi}}

\institute { IFAE, Edifici Cn., Campus UAB, E-08193 Bellaterra, Spain
 \and INAF National Institute for Astrophysics, I-00136 Rome, Italy
 \and Universit\`a  di Siena, and INFN Pisa, I-53100 Siena, Italy
 \and Universidad Complutense, E-28040 Madrid, Spain
 \and Technische Universit\"at Dortmund, D-44221 Dortmund, Germany
 \and Max-Planck-Institut f\"ur Physik, D-80805 M\"unchen, Germany
 \and University of \L\'od\'z, PL-90236 Lodz, Poland
 \and Inst. de Astrof\'{\i}sica de Canarias, E-38200 La Laguna, Tenerife, Spain
 \and Depto. de Astrof\'{\i}sica, Universidad de La Laguna, E-38206 La Laguna, Spain
 \and Deutsches Elektronen-Synchrotron (DESY), D-15738 Zeuthen, Germany
% ...
}

\date{Received ... / Accepted ... Draft version \today}

\abstract
{BL Lacertae objects are variable at all energy bands on time scales down to minutes. To construct and interpret their spectral energy distribution (SED), simultaneous broad-band observations are mandatory. Up to now, the number of objects studied during such campaigns is very limited and biased towards high flux states.}
{We present the results of a dedicated multi-wavelength study of the high-frequency peaked BL Lacertae (HBL) object and known TeV emitter 1ES\,2344$+$514 by means of a pre-organised campaign.}
{The observations were conducted during simultaneous visibility windows of MAGIC and AGILE in late 2008. The measurements were complemented by Mets\"ahovi, RATAN-600, KVA$+$Tuorla, \emph{Swift} and VLBA pointings. Additional coverage was provided by the ongoing long-term F-GAMMA and MOJAVE programs, the OVRO 40-m and CrAO telescopes as well as the \emph{Fermi} satellite. The obtained SEDs are modelled using a one-zone as well as a self-consistent two-zone synchrotron self-Compton model.}
{1ES\,2344$+$514 was found at very low flux states in both X-rays and very high energy gamma rays. Variability was detected in the low frequency radio and X-ray bands only, where for the latter a small flare was observed. The X-ray flare was possibly caused by shock acceleration characterised by similar cooling and acceleration time scales. \emph{MOJAVE VLBA monitoring reveals a static jet whose components are stable over time scales of eleven years, contrary to previous findings}. There appears to be no significant correlation between the 15\,GHz and R-band monitoring light curves. The observations presented here constitute the first multi-wavelength campaign on 1ES\,2344$+$514 from radio to VHE energies and one of the few simultaneous SEDs during low activity states. The quasi-simultaneous \emph{Fermi}-LAT data poses some challenges for SED modelling, but in general the SEDs are described well by both applied models. The resulting parameters are typical for TeV emitting HBLs. Consequently it remains unclear whether a so-called quiescent state was found in this campaign.}
{}


\titlerunning{The Simultaneous Low State SED of 1ES\,2344$+$514 from Radio to VHE}
\maketitle

\section{Introduction}\label{intro}
The number of known extragalactic very high energy (VHE, $\ga$\,100\,GeV) gamma-ray sources has been increasing steadily in the past seven years and now exceeds 50 (November 2012)\footnote{\href{http://tevcat.uchicago.edu/}{http://tevcat.uchicago.edu/}}. Most of these sources are X-ray bright BL Lacertae (BL Lac) objects. In BL Lacs the relativistic jet is nearly aligned with the line of sight and the resulting large relativistic beaming causes rapid variability in all energy regimes from radio wavelengths to VHE gamma rays. The spectral energy distribution (SED) of these objects shows two peaks; the low energy peak is attributed to synchrotron emission, emitted by relativistic electrons spiralling in the magnetic field lines of the jet, while the high energy peak is generally considered to be produced by inverse Compton scattering. The seed photons for the Compton scattering can be the synchrotron photons themselves \citep[synchrotron self Compton, SSC, e.g.][]{Maraschi92,Bloom96} or photons from an external radiation field \citep[accretion disk, broad line region clouds or infrared torus;][]{Dermer93,Sikora94,Blazejowski00}. An alternate source has been proposed, that the gamma rays are produced by hadronic processes, that is by proton initiated cascades or directly through proton synchrotron radiation \citep{Mannheim92,Muecke03}.

BL Lac objects were historically divided into two subclasses, depending on the energy of the synchrotron peak. The class boundaries can be loosely defined such that low energy peaking BL Lac objects (LBLs) have their peak at $10^{14-15}$\,Hz (optical regime) and high energy peaking BL Lacs (HBLs) at $> 10^{15}$\,Hz (UV to hard X-rays) \citep[e.g.][]{Padovani95}. The class intermediate to these two was introduced by \citet{Laurent-Muehleisen99}, noting that BL Lacs exhibit a continuous range in SED peak energy rather than a dichotomy. The BL Lac sources detected in VHE gamma rays mostly belong to the HBL class. Their spectral energy distributions can be described with one-zone SSC emission, but the modelling requires rather high jet speeds while Very Long Baseline Interferometry (VLBI) observations have shown that the parsec-scale jets of these objects are comparably slow \citep[Lorentz factor $\Gamma_\mathrm{model} \approx 25$ compared to $\Gamma_\mathrm{VLBI} \lesssim 5$;][]{Piner10}. Therefore it has been suggested that the jet is decelerating \citep{Georganopoulos03} or has a spine-and-sheath structure \citep{Ghisellini05}. Recent VLBI observations of the electric vector position angle and fractional polarisation distribution in TeV blazars support the spine-sheath scenario \citep{Piner10}.

BL Lac objects show variability at all bands from radio to VHE gamma rays. The variability amplitudes vary between the different energy regimes and from source to source. The VHE gamma-ray detected X-ray selected BL Lacs are typically quite faint and mildly variable in the radio, show a large range of variability in the optical band and are strongly variable in X-rays. In the gamma-ray band they are often mildly variable at sub-GeV\,--\,GeV energies, while in VHE gamma rays some of the sources show extreme variability with amplitudes exceeding one order of magnitude and flux doubling time scales as short as minutes \citep[e.g.\ Mrk\,421, Mrk\,501, PKS\,2155$-$304;][]{VERITAS_Mrk421,MAGIC_Mrk501,2155_flare} whereas others vary with smaller amplitude \citep[e.g.\ 1ES\,1215$+$303, PG\,1553$+$113;][]{MAGIC1215,MAGIC1553}. The variability is typically described in terms of ``quiescence'' and ``flaring'' epochs \citep[e.g.][]{VERITAS_Mrk501_MWL}.

Due to their variability and their broad-band emission, the SED of BL Lacs has to be based on simultaneous observations at all energy ranges (simultaneous multi-wavelength [MW] campaigns). For many sources the observations are concentrated on flaring epochs due to a higher detection probability. Simultaneous MW observations from radio to VHE gamma rays in low flux states were for a long time scarce for these objects due to limited sensitivity of the first generation of gamma-ray instruments. Even today such observations are mostly available for the three brightest objects, Mrk\,421, Mrk\,501 and PKS\,2155$-$304 \citep[see e.g.\ the most recent campaigns in][]{Mrk501_MWL,Mrk421_MWL,2155_MWL}.

1ES\,2344$+$514 is an HBL at redshift $z = 0.044$ \citep{Perlman96}. It was first detected at VHE gamma rays (above 300\,GeV) by the Whipple telescope in 1995 during a flare with a flux $F \left(> 350\,\mathrm{GeV}\right) = \left(6.6 \pm 1.9\right) \cdot 10^{-11}\,\mathrm{ph}\,\mathrm{cm}^{-2}\,\mathrm{s}^{-1}$ \citep{discovery_2344} and was at that time only the third known extragalactic VHE gamma-ray source. Follow-up observations in a lower state did not result in detections with high statistical significance until the MAGIC observations in 2005\,--\,2006 \citep{MAGIC_2344}. The source was not seen by EGRET \citep[e.g.][]{Mukherjee97} but was detected by the \emph{Fermi}-LAT with a flux $F \left(1-100\,\mathrm{GeV}\right) = \left(1.55 \pm 0.18\right) \cdot 10^{-9}\,\mathrm{ph}\,\mathrm{cm}^{-2}\,\mathrm{s}^{-1}$ and a hard power law spectral index ($1.72 \pm 0.08$) as reported in the \emph{Fermi}-LAT Second Source Catalog \citep[2FGL;][]{Fermi_2FGL} (see also Sect.~\ref{chap:SED_results}). Like most HBLs it does not exhibit strong variability in the \emph{Fermi} band \citep[variability index $\sim$\,28 in 2FGL, while an index of $> 41$ was required to reject the null hypothesis of no variability at the 99\,\% confidence level;][]{Fermi_2FGL}. Note that 1ES\,2344$+$514 is formally not listed as a ``clean'' source in the \emph{Fermi} AGN Catalog due to its low Galactic latitude but nevertheless appears in the corresponding source tables.

In the X-ray band the source is bright with a 2\,keV flux density of 1.14\,$\mu$Jy \citep{Perlman96} and showed strong spectral variability with the synchrotron peak shifting to higher energies with increasing flux \citep{Giommi00}. In the high state, the synchrotron peak frequency was at or above 10\,keV, making 1ES\,2344$+$514 one of the few so-called ``extreme blazars'' \citep{extreme_blazars} with synchrotron peak frequencies in the hard X-rays. \emph{Chandra} observations revealed diffuse X-ray emission as well as seven individual point sources in its environment \citep{Donato03}.

In the optical band the overall brightness of the source shows only very moderate variability (of the order of 0.1 mag). This is due to the bright host galaxy which contributes $\sim$\,90\,\% to the observed flux \citep{Nilsson07_2344}.

In the radio band the source is rather faint with a core flux density $S_\mathrm{core} \left(5\,\mathrm{GHz}\right) \approx 0.07$\,Jy measured by VLBI \citep{Giroletti04} and an overall flux density on arcsecond scales of $S_\mathrm{arcsec} \left(5\,\mathrm{GHz}\right) = \left(0.23 \pm 0.01\right)$\,Jy (average of 18 F-GAMMA\footnote{\href{http://www.mpifr-bonn.mpg.de/div/vlbi/fgamma/fgamma.html}{http://www.mpifr-bonn.mpg.de/div/vlbi/fgamma/fgamma.html}} single-dish observations from 02/2007 to 04/2009). Using Very Long Baseline Array (VLBA) imaging the apparent jet speeds of different components have been determined to be $\la 3\,c$ with the most robust measurement of $(0.62 \pm 0.05)\,c$ found for one individual feature \citep{Piner04,Piner10}. The lower frequency Very Large Array (VLA) maps (kpc scale) showed an extended and complex radio structure at 1.4\,GHz with $\sim$\,45$^\circ$ misalignment compared to higher frequency (5\,GHz, pc scale) radio maps \citep{Rector03,Giroletti04}.

\begin{figure}
\resizebox{\hsize}{!}{\includegraphics{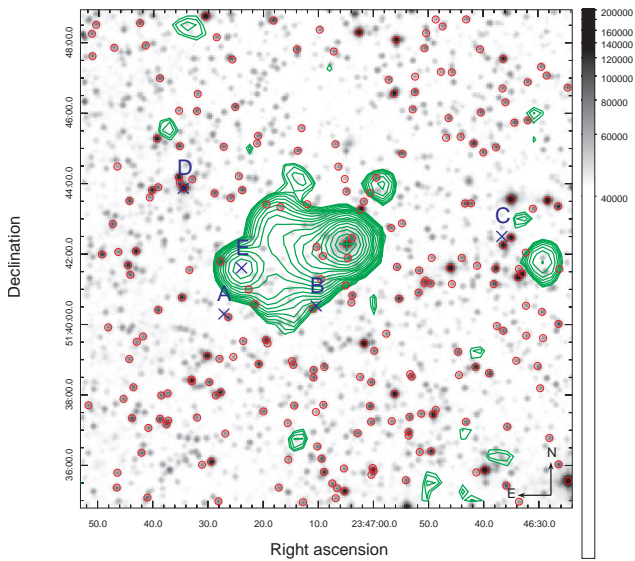}}
\caption{Sky map of the region around 1ES\,2344$+$514 (marked by a green cross). Green radio contours at 1.4\,GHz are overlaid on an R-band image. IR point sources are indicated by red circles. An ``X'' and its label mark individual components identified by \emph{Chandra} \citep{Donato03}. The logarithmic grey scale shows scaled densities. Radio contours are given from 0.001\,Jy/beam to 0.241\,Jy/beam in 20 logarithmically scaled steps. Only IR sources of J magnitude $< 15$ are displayed. Data reference: X-rays: \citet{Donato03}; optical: DSS2red; IR: 2MASS; radio: NVSS, obtained from NED.}\label{fig:2344_skymap}
\end{figure}

The combination of archival, non-simultaneous data in the radio, optical and X-ray regime reveals that only one of the individual X-ray components in the field of view of 1ES\,2344$+$514 is bright at 1.4\,GHz (component ``E'', see Fig.~\ref{fig:2344_skymap}). This component coincides very well with the radio feature reported by \citet{Rector03} and \citet{Giroletti04}, but is not present in the IR or R-band. Consequently, there are no other potential VHE candidate sources in the immediate vicinity of the source at an angular separation smaller than the MAGIC angular resolution of $\sim$\,0.1$^\circ$. The nature of the radio feature can not be identified unambiguously. Pulsars, being faint in the optical regime, would be viable candidates. However, \citet{Giroletti04} found a connection of the emission between the feature and the core in VLA radio images. Also the proximity between these two (angular distance of $\sim$\,180\arcsec, i.e.\ $\sim$\,160\,kpc) indicates that they might be related. The jet of the AGN may bend on kpc scales by $\sim$\,45$^\circ$ and interact with the intergalactic medium, resulting in a radio hot spot. The wide opening angle of the jet and the low surface brightness on these scales do not support the interpretation of the feature as a hot spot at this distance from the core though. Moreover, this would be in contradiction to the unification scenario where the BL Lacs are suggested to be beamed FR-I radio galaxies \citep{Urry95}. Note, however, that similar results have been found by e.g.\ \citet{Landt08,Kharb10}. Future VLBI measurements of the radio spectrum of the feature may distinguish between the radio hot spot or foreground/background source interpretation.

To date, 1ES\,2344$+$514 has been studied in only one MW campaign that included gamma-ray observations, conducted by \emph{RXTE}, \emph{Swift} and VERITAS \citep{VERITAS_2344_MWL}. \citet{Giommi11} reported \emph{Planck}, \emph{Swift} and \emph{Fermi} observations, covering energies from radio to GeV, but detecting the source only in the UV and X-ray bands. In this paper we present the first simultaneous radio to VHE gamma-ray observations of 1ES\,2344$+$514. The campaign was organised independently of the flux state to allow investigations of a low, possibly ``quiescent'', state of the source. The observations were scheduled to give the best simultaneous coverage between the different instruments, with less than a day time difference between VHE, X-ray and optical bands. The time delays with respect to radio observations were longer due to the longer variability time scale in this energy regime. The campaign took place in late 2008 shortly after the launch of the \emph{Fermi} satellite. In total six radio observatories contributed, including VLBA imaging of the source in several frequency bands. 1ES\,2344$+$514 was monitored in the optical R-band by the CrAO, KVA and Tuorla telescopes, in ultraviolet and X-rays by \emph{Swift} UVOT and XRT and in high energy (HE) gamma rays by AGILE and \emph{Fermi}. The core part of the campaign was centred around the MAGIC VHE gamma-ray observations of the source. Parts of the MW data sets have been presented in \citet{RuegamerICRC11,RuegamerFermi11}. In this paper we present the complete results of the campaign. We adopt a cosmology with $\Omega_\mathrm{m} = 0.27$, $\Omega_\Lambda = 0.73$ and $H_0 = 71\;\mathrm{km}\;\mathrm{s}^{-1}\;\mathrm{Mpc}^{-1}$ for calculating radio component linear sizes and proper motions.

The paper is organised as follows: in Sect.\ 2, we present short descriptions of the various participating instruments, their observations as well as the corresponding data analyses. The results will be shown in Sect.\ 3 and discussed in Sect.\ 4 including the spectral energy distributions and the theoretical models. Final remarks are given in Sect.\ 5.

\section{Instruments, Multi-Wavelength Observations and Data Analysis}\label{chap:observations}
In this section, the instruments participating in the MW campaign, their observations and data reduction processes will be presented ordered by their wavelength regime. A summary of the observation dates is given in Table~\ref{tab:observations}.

\begin{table}
\caption{Multi-wavelength observations of 1ES\,2344$+$514.}\label{tab:observations}
\centering
\begin{tabular}{llll}
\hline\hline
Instrument & Band\tablefootmark{a} & Observation Date\tablefootmark{b}\\\hline
Effelsberg & radio & 56; 78; 106; 155\\
IRAM & radio & 25; 106; 137; 178\\
Mets\"ahovi & radio & 30; 46; 124; 127; 130; 136; 138\\
OVRO & radio & 61\,--\,179\\
RATAN-600 & radio & 29\,--\,42\\
VLBA & radio & 61\,--\,62\\
CrAO & R-band & 74; 77; 85; 101; 105; 112; 117\\
KVA$+$Tuorla & R-band & 22\,--\,134\\
\emph{Swift} & UV \& X-rays & 30; 45\,--\,84\\
AGILE & HE gamma rays & 70\,--\,100\\
\emph{Fermi} & HE gamma rays & 59\,--\,100\\
MAGIC-I & VHE gamma rays & 59\,--\,100\\\hline
\end{tabular}
\tablefoot{
\tablefoottext{a}{The exact energy bands are given in Sect.~\ref{chap:observations}.}
\tablefoottext{b}{The dates are given in MJD$-$54700 and rounded down. In the case of OVRO, RATAN-600, KVA$+$Tuorla, \emph{Swift} XRT and MAGIC-I, the given observation periods were not covered continuously.}
}
\end{table}

\subsection{The MAGIC Telescope}
The MAGIC (Major Atmospheric Gamma-ray Imaging Cherenkov) project operates a system of two 17-m Imaging Air Cherenkov Telescopes located on the Canary Island of La Palma 2200\,m above sea level \citep{MAGIC_Crab_stereo}. MAGIC  has been operating in stereoscopic mode since 2009, accordingly the observations presented in this paper were conducted with MAGIC-I only (mono mode). MAGIC-I had a standard trigger threshold of 60\,GeV for observations at low zenith angles, an angular resolution of $\sim$\,0.1$^\circ$ for single events and an energy resolution above 150\,GeV of $\sim$\,25\,\% \citep[for details, see][]{MAGIC_Crab}.

MAGIC-I observed 1ES\,2344$+$514 from 20/10/2008 to 30/11/2008 at zenith angles between 23$^\circ$ and 31$^\circ$ for a total of 26.4\,hours in so-called wobble mode, where the source was displaced by 0.4$^\circ$ from the camera centre in order to allow the recording of simultaneous OFF-source data with the same offset from the camera centre \citep{Daum97}.

The data were analysed as described in \citet{MAGIC_Mrk421_MWL} with the exception of the signal arrival time extraction. Instead of determining the arrival time of the signal at the pulse maximum, which was needed at that time due to the special nature of those data, the standard method of determining the signal arrival time at half of the rising flank was used here. 20.8\,hours of data survived the quality selection. Background suppression was accomplished by a cut in shower area versus shower SIZE (i.e.\ total photoelectron content), optimised on 0.7\,hours of data from a high state of Mrk\,421 taken during the same observing period as 1ES\,2344$+$514 and hence with similar data quality and observation conditions. The significance of the signal was determined by a cut in $\theta^2$ optimised also on the Mrk\,421 data set, where $\theta$ is the angular distance between the expected and reconstructed source position. All significances of the VHE signals given in the following sections were determined by Eq.~17 of \citet{LiMa83} with $\alpha = 1/3$, i.e.\ using 3 OFF regions.

The source spectrum has been derived from events with $\theta^2 < 0.046$\,deg$^2$, yielding an analysis threshold of $\sim$\,190\,GeV. Upper limits (UL) were calculated by applying model 4 of \citet{Rolke05} using a confidence level (c.l.) of 95\,\%. The conversion from the differential spectrum to spectral energy density $\nu F_{\nu}$ has been accomplished by multiplying the differential flux with the energy of the Lafferty-Wyatt bin centre \citep{LaffertyWyatt95} squared.

The MAGIC analysis results presented here were confirmed by an independent internal analysis.

\subsection{The AGILE Satellite}
AGILE (Astrorivelatore Gamma ad Immagini LEggero) \citep{AGILE2} is a scientific mission of the Italian Space Agency dedicated to the observation of astrophysical sources of high energy gamma rays in the 30\,MeV\,--\,50\,GeV energy range, with simultaneous X-ray imaging capability in the 18\,--\,60\,keV band. AGILE is the first high-energy mission which makes use of a silicon detector for the gamma ray to pair conversion. The AGILE payload combines for the first time two coaxial instruments: the Gamma-Ray Imaging Detector (GRID, composed of a 12-planes Silicon-Tungsten tracker, a Cesium-Iodide mini-calorimeter and the anti-coincidence shield) and the hard X-ray detector Super-AGILE. The use of the silicon technology provides good performance of the gamma ray GRID imager in a relatively small and compact instrument: an effective area of the order of 500\,cm$^2$ at several hundred MeV, an angular resolution of around 3.5$^\circ$ at 100\,MeV, decreasing below 1$^\circ$ above 1\,GeV, a very large field of view ($\sim$\,2.5\,sr) as well as accurate timing, positional and attitude information.

During the period 07/2007\,--\,10/2009, AGILE was operated in ``pointing observing mode'', characterised by long observations called Observation Blocks (OBs), typically of two to four weeks duration, mostly concentrated along the Galactic plane. Since 11/2009 the satellite has been operating in ``spinning observing mode'', surveying a large fraction (about 70\,\%) of the sky each day. The time period covered by the 2008 MW campaign includes the AGILE OB 6400, publicly available from the  ASDC Multimission Archive web page\footnote{\href{http://www.asdc.asi.it/mmia/index.php?mission=agilemmia}{http://www.asdc.asi.it/mmia/index.php?mission=agilemmia}}. 1ES\,2344$+$514 was observed by AGILE at $\sim$\,40$^\circ$ off-axis from the mean pointing direction in the time window 31/10/2008 to 30/11/2008.

AGILE-GRID data from the official Processing Archive (SPINNING sw\,=\,5\_21\_18\_19 and POINTING sw\,=\,5\_19\_18\_17), obtained by using the AGILE Standard Analysis Pipeline \citep{AGILE1cat}, were analysed using the latest scientific software (AGILE\_SW\_5.0\_SourceCode) and in-flight calibrations (I0023) publicly available since 30/09/2011 at the ASDC site\footnote{\href{http://agile.asdc.asi.it/publicsoftware.html}{http://agile.asdc.asi.it/publicsoftware.html}}. Counts, exposure, and Galactic background gamma-ray maps were created with a bin-size of 0.3$^\circ \times 0.3^\circ$, for $E > 100$\,MeV, selecting only events flagged as confirmed gamma-ray events. Events collected during passages of the South Atlantic Anomaly or whose reconstructed directions form angles with the satellite-Earth vector smaller than 90$^\circ$ were rejected to avoid Earth albedo contamination. In order to derive the estimated flux (or flux upper limits) of the source we ran the AGILE point source analysis software based on the maximum likelihood technique using a radius of 10$^\circ$.

\subsection{\emph{Fermi}-LAT}\label{chap:Intro_Fermi}
The \emph{Fermi} satellite started taking official science data on 4/08/2008 \citep{LAT_instrument}. Two different detectors are on board: the Gamma-ray Burst Monitor (GBM), sensitive at low energies (8\,keV\,--\,40\,MeV), and the Large Area Telescope (LAT), sensitive at high energies (20\,MeV\,--\,$> 300$\,GeV).

Typically, the \emph{Fermi} satellite is rocked first towards the north pole of the orbit and then, in the next orbit, towards the south, alternating in this way the pointing in every orbit. This main operating mode, called ``All-Sky scanning mode'', allows for full sky coverage every two orbits, or three hours.

The LAT is a large field of view ($\sim$\,2.4\,sr) electron-positron pair conversion telescope made up of a high-resolution silicon microstrip tracker, a CsI hodoscopic electromagnetic calorimeter and an anti-coincidence detector for the identification of charged particle backgrounds. The full description of the instrument and its performance can be found in \citet{LAT_instrument}. The LAT point spread function (PSF) depends strongly upon the energy of the impinging gamma ray and on the depth of the conversion point in the tracker, and mildly upon the incidence angle. For normal-incidence conversions in the upper section of the tracker, the PSF 68\,\% containment radius is 0.6$^{\circ}$ for 1\,GeV photons and amounts to $\sim$\,$0.04^\circ$ above 100\,GeV.

The \emph{Fermi}-LAT data for 1ES\,2344$+$514 presented here were obtained in the time period between 20/10/2008 22:35:00 UTC and 30/11/2008 21:31:00 UTC coordinated with the observations with MAGIC. The data have been analysed by using the standard \emph{Fermi}-LAT Science Tools software package, version 09-27-01 as described in the Cicerone website\footnote{\href{http://fermi.gsfc.nasa.gov/ssc/data/analysis/documentation/Cicerone/}{http://fermi.gsfc.nasa.gov/ssc/data/analysis/documentation/Cicerone/}}. The Pass 7 Source event class and P7SOURCE\_V6 instrument response functions \citep{LAT_instrument} were used in our analysis. We selected events in a region of interest (RoI) centred on the source position within $15^\circ$, having an energy between 100\,MeV and 300\,GeV. In order to avoid background contamination from the bright Earth limb, time intervals when the Earth entered the LAT RoI were excluded from the data set. In addition, events with zenith angles larger than $100^{\circ}$ with respect to the Earth reference frame \citep{Fermi_bright} were excluded from the analysis. The data were analysed with an \emph{unbinned} maximum likelihood technique, described in \citet{Mattox96}, using the analysis software ({\tt gtlike}) developed by the LAT team and described in the Cicerone website mentioned above. The fitting procedure maximises the likelihood acting simultaneously on the free spectral parameters for the source of interest, those of nearby gamma-ray sources and the diffuse backgrounds, modelled using \emph{ring\_2year\_P76\_v0} for the Galactic diffuse emission and \emph{isotrop\_2year\_P76\_source\_v0} for the extragalactic isotropic emission models\footnote{\href{http://fermi.gsfc.nasa.gov/ssc/data/access/lat/BackgroundModels.html}{http://fermi.gsfc.nasa.gov/ssc/data/access/lat/BackgroundModels.html}}. To maintain comparability, photon fluxes were converted to spectral energy densities applying the same method as used for AGILE.

In addition we also performed a dedicated analysis of the highest energy photons ($> 100$\,GeV) detected from 1ES\,2344$+$514 within the first 44\,months of LAT operation. Only events of the purest class (Pass\_7\_V6\_Ultraclean) from a 68\,\% containment radius around the direction of the source were considered for this analysis. \emph{Front} and \emph{back} photons, accordingly to the definition in \citet{LAT_instrument}, were treated separately, having a different distribution of the PSF. Since no results on such events over this long time scale have been reported in literature, the analysis has been applied to four additional TeV HBLs with a comparable redshift (Mrk\,421, Mrk\,501, Mrk\,180 and 1ES\,1959$+$650).

\subsection{Swift}\label{chap:Intro_Swift}
The \emph{Swift} satellite \citep{2004ApJ...611.1005G} is equipped with three telescopes, the Burst Alert Telescope \citep[BAT;][]{2005SSRv..120..143B} which covers the 14\,--\,195\,keV energy range, the X-ray telescope \citep[XRT;][]{2005SSRv..120..165B} covering the 0.2\,--\,10\,keV energy band, and the UV/Optical Telescope \citep[UVOT;][]{2005SSRv..120...95R} covering the 180\,--\,600\,nm wavelength range with V, B, U, UVW1, UVM2 and UVW2 filters.

\emph{Swift} XRT observed 1ES\,2344$+$514 from 09\,--\,11/2008 with a total of 21 exposures (see Table~\ref{tab:Swift_XRT_results}) with exposure times ranging from 200\,s to 5\,ks. The two exposures lasting well below 1\,ks were too short for deriving a flux and were therefore excluded from the analysis. The XRT data were processed with standard procedures using the FTOOLS task XRTPIPELINE (version 0.12.6) distributed by HEASARC within the HEASOFT package (v.6.10). Events with grades 0\,--\,12 were selected \citep[see][]{2005SSRv..120..165B} and latest response matrices available in the \emph{Swift} CALDB (v.20100802) were used. For the spectral analysis the source events were extracted in the 0.3\,--\,10\,keV range within a circle with a radius of 20 pixels ($\sim$\,47\arcsec). The background was extracted from an off-source circular region with a radius of 40 pixels. The spectra were extracted from the corresponding event files and binned using GRPPHA to ensure a minimum of 25 counts per energy bin, in order to guarantee reliable $\chi^2$ statistics \citep{Gehrels86}. Spectral analyses were performed using XSPEC version 12.6.0. The spectral index was determined using an absorbed power law fit ($f_0 \cdot E^{-\alpha} \cdot \mathrm{e}^{-\tau}$) from 0.3\,--\,10\,keV, with the absorption $\tau$ being the product of the absorption hydrogen-equivalent column density N$_\mathrm{H}$ and the element-specific energy-dependent photoelectric cross section $\sigma\left(E\right)$. N$_\mathrm{H}$ was fixed to the Galactic value in the direction of the source of $1.5 \cdot 10^{21}\,\mathrm{cm}^{-2}$ \citep{2005A&A...440..775K}. Not fixing this parameter, the XRT data analysis yields a value of $\left(2.0 \pm 0.2\right) \cdot 10^{21}\,\mathrm{cm}^{-2}$. Since some daily data sets showed hints of spectral curvature, also fits using a log-parabola model ($f_0 \cdot E^{-\left(a + b \log_{10}\left(E\right)\right)} \cdot \mathrm{e}^{-\tau}$) were performed. However, for the majority of the cases the log-parabola fit was not significantly preferred by a logarithmic likelihood ratio test over the simple power law model (see Table~\ref{tab:Swift_XRT_results}). Therefore, the simple power law results were used as a common basis.

For the long-term source evolution, 67 observations of 1ES\,2344$+$514 between 2005 and 2010 were analysed. A slightly different analysis procedure was used compared to the MW data reduction. The spectra were determined using XSELECT (V2.4b) to extract events with an energy of 0.5\,--\,10\,keV from the corresponding event files. The background was deduced from an annulus around the source with an inner radius of 50 pixels ($\sim$\,118\arcsec) and an outer radius of 70 pixels ($\sim$\,165\arcsec). Spectral analysis and binning was performed in ISIS (V 1.6.2-3), where a minimum signal to noise ratio of 5 was required for grouping the data. The spectral index was determined in the range 0.5\,--\,10\,keV using an absorbed power law fit. To calculate the integral flux the photon flux was evaluated on a fine grid between 2 and 10\,keV. The neutral hydrogen-equivalent column density was determined for each spectrum from the spectral fit, yielding for spectra with a d.o.f.\ $> 35$ a mean value of $\left(1.71 \pm 0.14\right) \cdot 10^{21}$\,cm$^{-2}$. Flux errors are given at a 90\,\% confidence level. The event counts for calculating the hardness ratios for the MW data were extracted applying this pipeline in the full energy range.

\emph{Swift} UVOT observed the source with all filters (V, B, U, UVW1, UVM2, UVW2) each time. The source counts were extracted from a circular region 5 arcsec-sized centred on the source position, while the background was extracted from a larger circular nearby source-free region. These data were processed with the {\tt uvotmaghist} task of the HEASOFT package. The observed magnitudes have been corrected for Galactic extinction $E_\mathrm{B-V} = 0.191$\,mag \citep{Schlafly11} using the extinction curve from \citet{Fitzpatrick99} adopting $R_\mathrm{V} = 3.07$ \citep{McCall00}. The host-galaxy flux contributes significantly to the observed brightness in the V-, B- and U-bands, however no values for the contribution were found in the literature. Therefore, the contribution is estimated from the R-band value from \citet{Nilsson07_2344} (aperture 5\arcsec) using the galaxy colours at $z = 0$ from \citet{Fukugita95} resulting in $\mathrm{V} = \left(1.96 \pm 0.16\right)$\,mJy, $\mathrm{B} = \left(0.95 \pm 0.20\right)$\,mJy and $\mathrm{U} = \left(0.22 \pm 0.20\right)$\,mJy. In these bands the host galaxy contributes $\sim$\,80\,--\,90\,\% to the measured flux and additionally the uncertainty of the host-galaxy contribution is rather large. Therefore these bands are not considered for spectral energy distribution modelling.

The magnitudes measured in the UV filters were converted to units of $\mathrm{erg}\,\mathrm{cm}^{-2}\,\mathrm{s}^{-1}$ using the photometric zero points as given in \citet{Breeveld11} and effective wavelengths and count-rate-to-flux ratios of GRBs from the \emph{Swift} UVOT CALDB 02 (v.20101130). \citet{Raiteri10} noted that these ratios are not necessarily applicable to BL Lac objects, due to their different spectrum and a B\,--\,V value typically larger than the applicable range. Therefore, they determined the UVOT effective wavelengths and count-rate-to-flux ratios anew (for BL Lacertae, an LBL at $z = 0.069$). We compare these values with the ones used in this work and find that the difference amounts to $\lesssim 1$\,\% for the V, B and U filters. In the case of the UV bands, the effective wavelengths (count-rate-to-flux ratios) are $\sim$\,7\,\% ($\sim$\,2\,\%), $\sim$\,3\,\% ($\sim$\,1\,\%) and $\sim$\,9\,\% ($\sim$\,13\,\%) larger for the UVW1, UVM2 and UVW2 filters, respectively. These differences are smaller than or comparable to the intrinsic errors of the corresponding values with the exception of the UVW2 count-rate-to-flux ratio (intrinsic error of $\sim$\,2\,\%). Therefore we did not apply a new calibration but increased the error of the UVW2 count-rate-to-flux ratio from $\sim$\,2\,\% to 13\,\% to account for a potential change in this value as large as found by \citet{Raiteri10}. However the actual uncertainty should be much below that, considering that some (if not most) of the difference between the ratios arises solely from using new effective wavelengths, which is not the case in this work.

\emph{Swift} BAT operates in full sky mode. The BAT data of 1ES\,2344$+$514, taken from the 58-Month Catalog\footnote{After an update, the 58-Month Catalog contains as of now (10/2012) the results from the first 66 months of observation.}, have been re-binned using the tool {\tt rebingausslc} from the HEASOFT package to weekly (7\,days), monthly (30.44\,days), quarterly (91.31\,days) and yearly (365.24\,days) bins. The default settings for the bin centre of {\tt rebingausslc} have been used, no trials have thus been made for selecting the binning. Integral fluxes were calculated according to \citet{BAT22} by multiplying the Crab-normalised count rate of 1ES\,2344$+$514 with the Crab flux measured in the same time interval and energy band. These fluxes were then converted to spectral energy densities in each energy band at the Lafferty-Wyatt bin position \citep{LaffertyWyatt95} assuming a simple power law with a spectral index of 2.62 as given in the BAT 58-Month Catalog.

\subsection{KVA and Tuorla}
1ES\,2344$+$514 has been monitored in the optical R-band by the Tuorla Blazar Monitoring Program since 2002\footnote{\href{http://users.utu.fi/kani/1m/}{http://users.utu.fi/kani/1m/}}. The observations are done using the Tuorla 1-m telescope (Finland) and the Kungliga Vetenskapsakademien (KVA) 35-cm telescope (La Palma). The latter can be controlled remotely from the Tuorla Observatory. In the following, ``KVA'' will be used as a synonym for ``KVA$+$Tuorla''. The source is typically observed a few times per week, but during the \emph{Swift} pointings mechanical problems prevented KVA observations. The photometric measurements are made in differential mode, i.e.\ by obtaining CCD images of the target and calibrated comparison stars in the same field of view \citep{Fiorucci98}. The magnitudes of the source and comparison stars are measured using aperture photometry and the (colour corrected) zero point of the image determined from the comparison star magnitudes. The object magnitude is computed using the zero point and a filter-dependent colour correction. Magnitudes are then transferred to linear flux densities using the formula $F = F_0 \cdot 10^{\mathrm{mag}/-2.5}$, where mag is the magnitude of the object and $F_0$ is a filter-dependent zero point \citep[in the R-band the value $F_0 = 3080$\,Jy is used from][]{Bessell79}.

Since 1ES\,2344$+$514 has a bright host galaxy and a nearby star that contributes to the observed flux, these contributions have to be removed in order to derive the core flux for the spectral energy distribution. \citet{Nilsson07_2344} determined these contributions which depend on seeing and the aperture used for the measurement. Since all observations for this campaign were done with constant aperture (7.5\arcsec) and in similar seeing conditions, we subtract a constant value of $\left(3.70 \pm 0.05\right)$\,mJy.

\subsection{CrAO}
Observations from the Crimean Astrophysical Observatory (CrAO) were obtained with the AZT-11 telescope and an FLI IMG1001E CCD camera, through an
R-band filter. Differential photometry was performed between the blazar and published comparison stars on the same CCD frame. The comparison stars and apertures used were the same as for KVA. The resulting magnitudes were converted to mJy using the standard formula. The CrAO flux densities were found to be $\sim$\,12\,\% lower than the KVA points and were shifted by a fixed value ($\sim$\,0.49\,mJy) to match the KVA observations. The corresponding shift has been deduced from the average flux density difference between both telescopes for nights with an observation time difference $< 0.3$\,days. Two out of seven data point pairs satisfied this condition. A difference of $\sim$\,10\,\% is expected due to CrAO using the Johnson R-band filter whereas KVA is measuring in the Cousins R-band filter.

\subsection{Effelsberg 100-m and IRAM 30-m Radio Telescopes}
Quasi-simultaneous cm-to-mm radio spectra have been obtained within the framework of a \emph{Fermi} related monitoring program of gamma-ray blazars, namely the F-GAMMA program \citep{Fuhrmann07, Angelakis08}. The total frequency range spans from 2.64\,GHz to 228.4\,GHz using the Effelsberg 100-m and IRAM 30-m telescopes. The millimetre observations are closely coordinated with the more general flux monitoring conducted by IRAM, and observations of both programs are included in this paper. 1ES\,2344$+$514 has been observed in late 2008 once a month with these facilities.

The Effelsberg measurements were conducted with the secondary focus heterodyne receivers at 2.64, 4.85, 8.35, 10.45, 14.60, 23.05, 32.00 and 43.00\,GHz. The observations were performed quasi-simultaneously with ``cross-scans'' (that is, slewing over the source position in azimuth and elevation direction), with an adaptive number of sub-scans for reaching the desired sensitivity \citep[for details see][]{Fuhrmann08, Angelakis08}. Subsequently, pointing off-set corrections, gain corrections and atmospheric opacity corrections have been applied to the data. The conversion to Jy has been done using the standard calibrators: 3C\,48, 3C\,161, 3C\,286, 3C\,295 and NGC\,7027. The standard deviation of the flux calibrators amounts to $<5$\,\% at 43.00\,GHz and $<1$\,\% at 2.64 GHz. The Effelsberg error bars are given including systematic uncertainties.

IRAM (Institut de Radioastronomie Millim\'{e}trique) operates a 30-m radio telescope located on Pico Veleta near Granada in Spain. The IRAM observations of 1ES\,2344$+$514 and primary/secondary calibrators were carried out with calibrated cross-scans using the receivers operating at 86.2 and 142.3\,GHz, occasionally also at 228.4\,GHz. The opacity corrected scans were converted into the standard temperature scale and finally corrected for small remaining pointing offsets and systematic gain-elevation effects. The conversion to the Jy flux density scale was done using the instantaneous conversion factors derived from the frequently observed primary (Mars, Uranus) and secondary (W3(OH), K3-50A, NGC\,7027) calibrators.

\subsection{Mets\"ahovi 14-m Radio Telescope}
The 37\,GHz observations were conducted with the 13.7-m diameter Mets\"ahovi radio telescope, which is a radome-enclosed paraboloid antenna in Finland. The measurements were made with a 1\,GHz-band dual beam receiver centred at 36.8\,GHz. The HEMPT (high electron mobility pseudomorphic transistor) front end operates at room temperature. The observations are ON\,--\,ON observations, alternating the source and the sky in each feed horn. A typical integration time to obtain one flux density data point is between 1200 and 1400\,s. The detection limit of the telescope at 37\,GHz is of the order of 0.2\,Jy under optimal conditions. Data points with a signal to noise ratio $< 4$ are considered as non-detections.

The flux density scale is set by observations of DR 21. The sources NGC\,7027, 3C\,274 and 3C\,84 are used as secondary calibrators. A detailed description of the data reduction and analysis is given in \citet{Metsaehovi}. The error estimate in the flux density includes the contribution from the measurement rms and the uncertainty of the absolute calibration.

\subsection{OVRO 40-m Radio Telescope}
Regular 15.0\,GHz observations of 1ES\,2344$+$514 were carried out as part of a high-cadence gamma-ray blazar monitoring program using the Owens Valley Radio Observatory (OVRO) 40-m telescope \citep{Richards11}. This program, which commenced in late 2007, now includes about 1600 sources, each observed with a nominal twice per week cadence. Data during the beginning of this MW campaign were unavailable due to a hardware outage. The OVRO 40-m results used in this paper span the period 22/10/2008 to 11/02/2012.

The OVRO 40-m uses off-axis dual-beam optics and a cryogenic high electron mobility transistor (HEMT) low-noise amplifier with a 15.0\,GHz centre frequency and 3\,GHz bandwidth. The total system noise temperature is about 52\,K, including receiver, atmosphere, ground, and CMB contributions. The two sky beams are Dicke-switched using the off-source beam as a reference, and the source is alternated between the two beams in an ON\,--\,ON fashion to remove atmospheric and ground contamination. A noise level of approximately 3\,--\,4\,mJy in quadrature with about 2\,\% additional uncertainty, mostly due to pointing errors, is achieved in a 70\,s integration period. Calibration is achieved using a temperature-stable diode noise source to remove receiver gain drifts. The flux density scale is derived from observations of 3C\,286 assuming a value of 3.44\,Jy at 15.0\,GHz \citep{Baars77}. The systematic uncertainty of about 5\,\% in the flux density scale is not included in the error bars. Complete details of the reduction and calibration procedure are found in \citet{Richards11}.

\subsection{RATAN-600}
The radio spectrum of 1ES\,2344$+$514 was observed with the 600-m ring radio telescope RATAN-600 \citep{RATAN} of the Special Astrophysical Observatory, Russian Academy of Sciences, located in Zelenchukskaya, Russia, from 20/09/2008 to 03/10/2008. The continuum spectrum was measured six times quasi-simultaneously (within several minutes) in a transit mode with six different receivers at the following central frequencies (and frequency bandwidths): 0.95\,GHz (0.03\,GHz), 2.3\,GHz  (0.25\,GHz), 4.8\,GHz (0.6\,GHz), 7.7\,GHz (1.0\,GHz), 11.2\,GHz (1.4\,GHz), 21.7\,GHz (2.5\,GHz). Due to radio frequency interference, we were unable to detect the source at the two longest wavelengths. An average spectrum of the six independent 5\,--\,22\,GHz measurements is presented in this paper. Details on the method of observation, data processing, and amplitude calibration are described by \citet{RATAN_analysis}. The data were collected using the southern sector with the Flat reflector.

\subsection{VLBA}
1ES\,2344$+$514 was observed with the VLBA \citep{Napier95} on 23/10/2008 at 4.6, 5.0, 8.1, 8.4, 15.4, 23.8 and 43.2\,GHz in the framework of a survey of parsec-scale radio spectra of 20 gamma-ray bright blazars \citep{Sokolovsky_magnetic_field}. The observations were conducted with ten on-source scans (each four to seven minutes long depending on frequency) spread over eleven hours. The data reduction was performed in the standard manner using the AIPS package \citep{Greisen90}. An amplitude calibration procedure similar to the one described in \citet{Sokolovsky_synchrotron} was applied, resulting in $\sim$\,5\,\% calibration accuracy at the 4.6\,--\,15.4\,GHz range and $\sim$\,10\,\% accuracy at 23.8 and 43.2\,GHz. The Difmap software \citep{Shepherd97} was used for imaging and modelling of the visibility ($uv$) data. The integrated parsec-scale flux densities were derived by summing all CLEAN \citep{Hoegbom74} components used to represent calibrated visibilities.

1ES\,2344$+$514 was also observed with VLBA at 15.4\,GHz during the campaign as a part of the MOJAVE (Monitoring Of Jets in Active galactic nuclei with VLBA Experiments)\footnote{\href{http://www.physics.purdue.edu/MOJAVE/}{http://www.physics.purdue.edu/MOJAVE/}} long-term program to monitor radio brightness and polarisation variations in jets associated with active galaxies visible in the northern sky. The data were analysed using the standard procedures \citep[see][]{Lister09_V,Lister09_VI}. Elliptical Gaussian components were used to determine positions and flux densities of individual emission regions within the source. The MOJAVE archive contains two sets of VLBA data on this source at 15.4\,GHz. One set contains four epochs published in \citet{Piner04} that span the range 10/1999 to 03/2000. The second consists of ten epochs covering 05/2008 to 11/2010.

\section{Results}\label{results}
\subsection{Very High Energy Gamma Rays}\label{chap:Results_VHE}
\begin{figure}
\resizebox{\hsize}{!}{\includegraphics{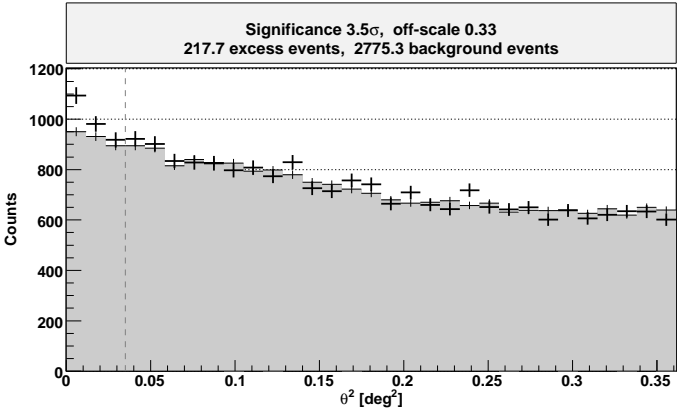}}
\caption{$\theta^2$ plot for 1ES\,2344$+$514. The vertical dashed line gives the $\theta^2$ cut, defining the signal region. The background events are shown by the grey area.}\label{fig:2344_thetasq}
\end{figure}

The MAGIC data analysis yielded a marginal signal of 3.5\,$\sigma$ for the complete data set (see Fig.~\ref{fig:2344_thetasq} and for detailed results Table~\ref{tab:MAGIC_results}), which is below the 5\,$\sigma$ standard for source discoveries in VHE astronomy. Since 1ES\,2344$+$514 is a well-established VHE emitter and the direct environment is lacking suitable alternative source candidates (see Sect.\ref{intro}), we assume that the entire excess comes from the source. The rather long observation time of $\sim$\,20\,hours and the fairly large events statistics not dominated by individual features in time makes us confident about the reliability of the signal. Therefore, we derived an average spectrum.

\begin{figure}
\resizebox{\hsize}{!}{\includegraphics{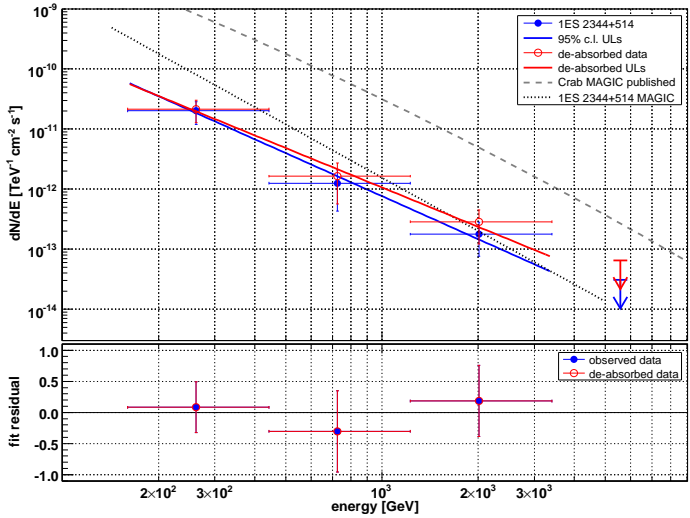}}
\caption{\emph{Top panel:} Measured (blue, filled circles) and de-absorbed (red, open circles) MAGIC spectra for 1ES\,2344$+$514, shown together with the MAGIC 2005 spectrum of 1ES\,2344$+$514 \citep{MAGIC_2344} as well as the MAGIC Crab spectrum \citep{MAGIC_Crab}. Upper limits were derived with a c.l.\ of 95\,\%. \emph{Bottom panel:} Fit residuals defined as the difference between flux and fit divided by the flux value.}\label{fig:2344_MAGIC_spectrum}
\end{figure}

The measured (EBL de-absorbed) spectra are rather well fitted ($\chi^2\mathrm{/d.o.f.} = 0.36/1$ for both of them; see also the residuals shown in Fig.~\ref{fig:2344_MAGIC_spectrum}) by a simple power law of the form
 \begin{equation}
\frac{\mathrm{d}N}{\mathrm{d}E} = f_0 \cdot 10^{-12}\,\mathrm{TeV}^{-1}\,\mathrm{cm}^{-2}\,\mathrm{s}^{-1} \cdot \left(E / E_0\right)^{-\alpha}\label{eq:PL}
\end{equation}
yielding $f_0 = 4.0 \pm 1.2$ ($4.8 \pm 1.5$) at $E_0 = 0.5$\,TeV and $\alpha = 2.4 \pm 0.4$ ($2.2 \pm 0.4$) (see Fig.~\ref{fig:2344_MAGIC_spectrum}). The given errors are statistical only. We adopt the MAGIC standard systematic errors of 16\,\% on the energy scale, 11\,\% on the flux normalisation and $\pm 0.2$ on the spectral index \citep{MAGIC_Crab}. The low redshift of the source renders differences between the current extragalactic background light (EBL) models negligible. Here, the effects of EBL absorption were corrected by applying the Kneiske ``lower limit'' model \citep{Kneiskelow}.

\begin{figure}
\resizebox{\hsize}{!}{\includegraphics{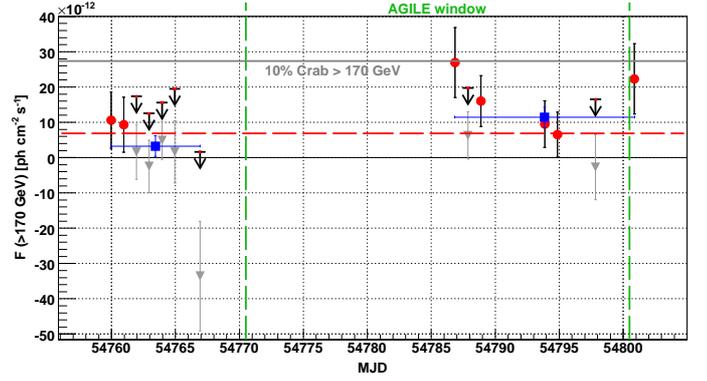}}
\caption{MAGIC light curve of 1ES\,2344$+$514 derived from this campaign. The red points give daily fluxes. For points having a flux consistent with or below zero, shown by the grey triangles, 95\,\% c.l.\ upper limits were calculated. The red bar on top of each upper limit arrow visualises the bin width. A fit with a constant to all daily flux points is shown by the red dashed line. Additionally, the fluxes for the two observation periods (see also Table~\ref{tab:MAGIC_results}) are given as blue full squares.}\label{fig:2344_MAGIC_LC}
\end{figure}

No significant variability could be found over the entire observation period on daily time scales, as can be seen from the light curve in Fig.~\ref{fig:2344_MAGIC_LC} (note that the fluxes are calculated subtracting OFF data from ON data and can therefore become negative) and the flux values given in Table~\ref{tab:MAGIC_results}. The overall flux $F \left(> 170\,\mathrm{GeV}\right)$ amounted to $(7.4 \pm 2.1) \cdot 10^{-12}\,\mathrm{ph}\,\mathrm{cm}^{-2}\,\mathrm{s}^{-1}$. A fit with a constant yields a $\chi^2\mathrm{/d.o.f.} = 19.0/13$, which gives $\sim$\,12\,\% probability for a constant flux. The low probability arises dominantly from the negative fluctuation around MJD 54767 and the highest flux point at MJD 54787. The latter is indicating a higher state of the source, but since the point is less than 2\,$\sigma$ above the fit line, it statistically does not give evidence for variability. The measurements exclude a rise in flux by more than a factor of $\sim$\,9 of the mean flux (derived from the highest 3\,$\sigma$ UL calculated for all light curve points), while the peak flux above 300\,GeV reported by \citet{VERITAS_2344_MWL} was a factor of $\sim$\,20 higher than the average flux $F \left(> 300\,\mathrm{GeV}\right) = \left(3.4 \pm 1.0\right) \cdot 10^{-12}\,\mathrm{ph}\,\mathrm{cm}^{-2}\,\mathrm{s}^{-1}$ found here. Fitting the period-wise light curve, the $\chi^2\mathrm{/d.o.f.} = 4.0/1$ (probability $\sim$\,5\,\%), which is still consistent with the hypothesis of a constant flux of the source.

Above 200\,GeV, the integral flux amounted to $\left(5.5 \pm 1.7\right) \cdot 10^{-12}\,\mathrm{ph}\,\mathrm{cm}^{-2}\,\mathrm{s}^{-1}$, more than a factor 4 lower than the former MAGIC detection (which, at the time, constituted the lowest flux measured of this source at VHE). Compared to the average flux measured by VERITAS $> 300$\,GeV in 2008 \citep[see][]{VERITAS_2344_MWL}, the average flux found here is still lower by a factor of $> 3$ and hence represents the lowest flux reported from 1ES\,2344$+$514 at low VHE thresholds up to now. At high energies the HEGRA collaboration reported a flux $F \left(> 970\,\mathrm{GeV}\right) = \left(6.0 \pm 1.9\right) \cdot 10^{-13}\,\mathrm{ph}\,\mathrm{cm}^{-2}\,\mathrm{s}^{-1}$ after 72.5\,hours of observation time between 1997 and 2002 \citep{2344_HEGRA}, which is comparable to our result ($F \left(> 970\,\mathrm{GeV}\right) = \left(4.8 \pm 3.1\right) \cdot 10^{-13}\,\mathrm{ph}\,\mathrm{cm}^{-2}\,\mathrm{s}^{-1}$).

Previous observations of 1ES\,2344$+$514 at VHE confirmed spectral variability, as expected for a BL Lac type object. The spectral index has ranged from $2.43 \pm 0.22_\mathrm{stat} \pm 0.15_\mathrm{syst}$ \citep{VERITAS_2344_MWL} to $2.95 \pm 0.12_\mathrm{stat} \pm 0.2_\mathrm{syst}$ \citep{MAGIC_2344} with a trend of a hardening of the spectrum with increasing flux. In contrast, the value of $2.4 \pm 0.4$ found here indicates a hard spectrum despite a very low flux state. However, these results are still consistent with most of the archival measurements due to the large statistical errors. A hard spectral index would imply that the second SED peak was located at unusually high energies for that flux level, opposite to the spectral hardening trend observed for the best studied blazars \citep[e.g.\ Mrk\,421, Mrk\,501, PKS\,2155$-$304;][]{Fossati08,MAGIC_Mrk501,HESS_2155}.

\subsection{High Energy Gamma Rays}\label{chap:Results_HE}
AGILE-GRID did not detect the source. The AGILE maximum likelihood analysis using the latest in-flight calibrations yielded a 95\,\% c.l.\ UL on the flux above 100\,MeV of $3.7 \cdot 10^{-8}\,\mathrm{ph}\,\mathrm{cm}^{-2}\,\mathrm{s}^{-1}$ from an effective exposure of $\sim\,2.8 \cdot 10^8\,\mathrm{cm}^2\,\mathrm{s}$ for the MW observation period. Searching for short flares on time scales of seven as well as two days did not yield any detection. Also for the entire period from 07/2007 up to 01/2011, the source was not detected by AGILE. A 95\,\% c.l.\ UL on the flux $> 100\,\mathrm{MeV}$ of $2.7 \cdot 10^{-8}\,\mathrm{ph}\,\mathrm{cm}^{-2}\,\mathrm{s}^{-1}$ was derived, consistent with the 2FGL average flux above 100\,MeV which is about $0.9 \cdot 10^{-8}\,\mathrm{ph}\,\mathrm{cm}^{-2}\,\mathrm{s}^{-1}$. Given the non-detection of the source we adopt a ``standard'' spectral photon index of 2.1 for the likelihood analysis.

\emph{Fermi}-LAT did not detect 1ES\,2344$+$514 between 0.1 and 300\,GeV during the campaign (effective exposure: $\sim\,3.7 \cdot 10^9\,\mathrm{cm}^2\,\mathrm{s}$). The data were searched for short-time variability on daily and weekly time scales without a clear sign of such. The 2-year Catalog public light curve does not show significant variability on time scales of months around the time of the MW campaign. Upper limits at a 95\,\% c.l.\ have been determined applying the standard Bayesian approach for the MW time slot, assuming a spectral index of 2.1 to be consistent with the AGILE calculations. These amount to (in $\mathrm{ph}\,\mathrm{cm}^{-2}\,\mathrm{s}^{-1}$) $3.0 \cdot 10^{-8}$ (0.1\,--\,0.3\,GeV), $6.7 \cdot 10^{-9}$ (0.3\,--\,1.0\,GeV), $2.7 \cdot 10^{-9}$ (1.0\,--\,3.0\,GeV), $8.8 \cdot 10^{-10}$ (3.0\,--\,10\,GeV) and $8.6 \cdot 10^{-10}$  (10\,--\,100\,GeV).

1ES\,2344$+$514 is rather dim for a TeV AGN in the \emph{Fermi} band. It was detected for the first time after 5.5 months of observations \citep{Fermi_TeVselected}. From the first \citep[1FGL;][]{Fermi_1FGL} to the second \citep{Fermi_2FGL} LAT Source Catalog listing, the measured fluxes from 1\,--\,100\,GeV and spectral power law indices changed from $\left(1.40 \pm 0.30\right) \cdot 10^{-9}\,\mathrm{ph}\,\mathrm{cm}^{-2}\,\mathrm{s}^{-1}$ to  $\left(1.55 \pm 0.18\right) \cdot 10^{-9}\,\mathrm{ph}\,\mathrm{cm}^{-2}\,\mathrm{s}^{-1}$ and $1.57 \pm 0.12$ to $1.72 \pm 0.08$, respectively. These values are consistent within the statistical errors, indicating that the spectral shape did not change significantly on these time scales. Also the monthly light curve shows mostly upper limits and marginal detections without signs of major flares. In fact, only one flux point from the monthly binned \emph{Fermi}-LAT data is available for 1ES\,2344$+$514 within the first nine months of regular measurements, the remaining observations resulted in ULs.

Consequently, 1ES\,2344$+$514 seems to be, within the limits of the AGILE-GRID and \emph{Fermi}-LAT sensitivities, a rather stable and weak source in the HE gamma-ray band over long time scales. Hence, archival data should yield a fairly good estimate of the actual flux during this MW campaign. We therefore use the spectral information from 1FGL on a quasi-simultaneous basis for SED modelling (see Sect.~\ref{chap:SED}).

The LAT high energy analysis revealed nine events with energies in excess of 100\,GeV within the first 44 months of operation from 1ES\,2344$+$514, the highest energy photon having an energy of nearly 500\,GeV (see Table~\ref{tab:FermiHE}). We compare these with the number of events detected from four similar sources (see Sect.~\ref{chap:Intro_Fermi}) in Table~\ref{tab:FermiHE_comp}. An investigation of the distribution of event energies is strongly limited by the small event statistics, but judging from Fig.~\ref{fig:FermiHE_comp}, most of them are clustered for Mrk\,421 at 100\,GeV, whereas the distribution seems to be shifted to $\sim$\,150\,GeV for Mrk\,501 and $\sim$\,200\,GeV for 1ES\,2344$+$514. If real, distinct HE flares may be responsible for most of the events $> 100$\,GeV detected from Mrk\,501 and 1ES\,2344$+$514 (we note that the events are not clustered in time), in contrast to Mrk\,421 for which the distribution indicates a constantly high flux at HE.

The number of events should be correlated directly with the source luminosity. Determining the latter at 60\,GeV \citep[from their respective photon fluxes between 10 and 100\,GeV in][]{Fermi_2FGL} and normalising the photon counts to the distance of 1ES\,2344$+$514, a linear fit for the five sources yields the expected correlation with a slope of $\left(0.99 \pm 0.24\right)$ counts per $10^{43}$\,erg\,s$^{-1}$ (not shown). This indicates that the 2FGL fluxes are a suitable representation of the average source behaviour. The goodness of the linear fit is rather low though, having a $\chi^2\mathrm{/d.o.f.} = 7.4/3$, but is preferred by a logarithmic likelihood ratio test with 98.9\,\% over a fit with a constant ($\chi^2\mathrm{/d.o.f.} = 25.5/4$). This is a consequence of the comparably low number of counts from 1ES\,1959$+$650, which may arise from the flatter spectral index at HE, and the high number of events detected from 1ES\,2344$+$514 (which should be 2\,--\,3 according to its luminosity). Considering the similar luminosities of the sources, the reason should be a higher flaring duty cycle rather than a higher long-term average flux of the source, which would be in line with the interpretation of the observed shift in event energy distributions. Alternatively, the event counts may also be artificially increased by false identification of Galactic foreground events of 1ES\,2344$+$514, being located at a low Galactic latitude of $-9.9^\circ$. However, applying the same analysis to two regions containing no HE source at the same Galactic latitude as 1ES\,2344$+$514, but 2.5$^\circ$ away from the object, did not result in the detection of any event with energy $> 100$\,GeV.

The weakness of this investigation is the low statistical basis of only five sources. Additionally, we note that the events above 100\,GeV have been extracted from 44 months of observations, whereas the luminosities were determined from 2FGL (24 months). These arguments render our conclusions rather speculative. A catalogue of sources with events $> 100$\,GeV based on longer observation times is needed to conduct a more reliable study.

\begin{table}
\caption{\emph{Fermi}-LAT detected events with an energy $> 100$\,GeV within the first 44 months of operation from the direction of 1ES\,2344$+$514 (R.A.\ 356.77$^\circ$, Dec.\ 51.71$^\circ$).
}\label{tab:FermiHE}
\centering
\begin{tabular}{lllll}
\hline\hline
MJD & Energy\tablefootmark{a} & R.A.\tablefootmark{b}  & Dec.\tablefootmark{c} & Sep.\tablefootmark{d}\\
& [GeV] & [$^\circ$] & [$^\circ$] & [arcmin]\\\hline
54879.961 & 221 & 356.59 & 51.80 & \phantom{0}9\\
54992.961 & 174 & 356.79 & 51.61 & \phantom{0}6\\
55041.439 & 283 & 356.73 & 51.75 & \phantom{0}3\\
55358.826 & 495 & 356.86 & 51.63 & \phantom{0}6\\
55553.247 & 201 & 357.01 & 51.60 & 10\\
55896.009 & 114 & 356.79 & 51.68 & \phantom{0}2\\
55702.733 & 207 & 356.97 & 51.63 & \phantom{0}9\\
55936.262 & 107 & 356.79 & 51.91 & 11\\
55948.736 & 231 & 356.74 & 51.73 & \phantom{0}2\\\hline
\end{tabular}
\tablefoot{
\tablefoottext{a}{Energy,}
\tablefoottext{b}{right ascension (J2000) and}
\tablefoottext{c}{declination (J2000) of the event.}
\tablefoottext{d}{Angular separation between the event direction and 1ES\,2344$+$514.}
}
\end{table}

\subsection{X-Rays}\label{chap:Results_Swift}
\begin{figure*}
\resizebox{\hsize}{!}{\includegraphics{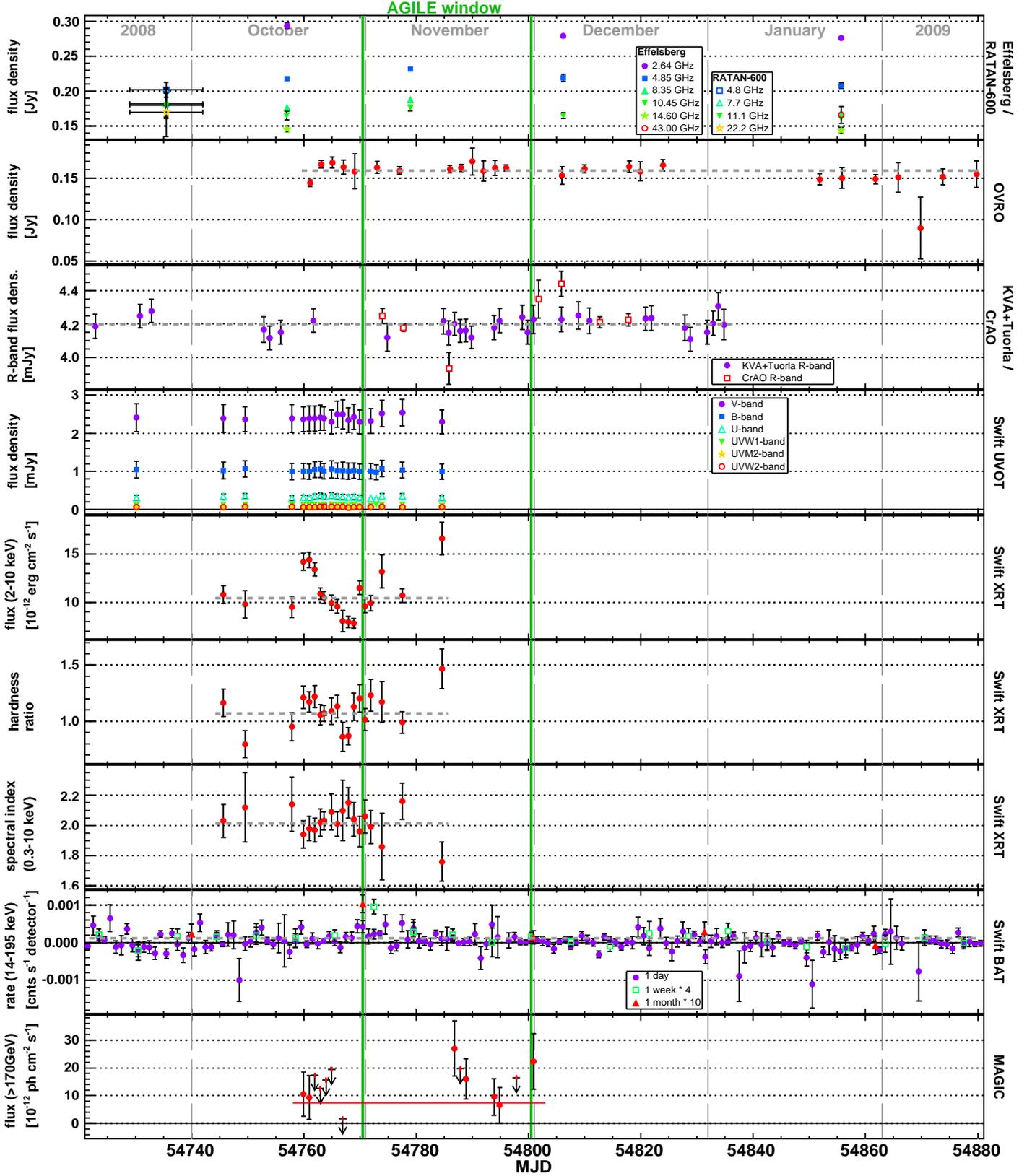}}
\caption{Combined 1ES\,2344$+$514 multi-wavelength light curve derived from this campaign. The observation window of AGILE is shown by the vertical solid green lines. Horizontal short-dashed lines represent fits with a constant to the data points. In the R-band, the fit is being applied to the combined KVA and CrAO data points. Hardness ratio is defined as the event counts between 2\,--\,10\,keV divided by the counts between 0.2\,--\,1\,keV. The \emph{Swift} BAT data are shown for several bin widths some of which have been multiplied by a factor for clarity as indicated in the panel legend. The fit shown in the BAT panel has been calculated for the 1 week binning. One point from the daily BAT light curve (MJD 54867.5, rate $-\left(6.08 \pm 3.49\right) \cdot 10^{-3}\,\mathrm{cnts}\,\mathrm{s}^{-1}\,\mathrm{detector}^{-1}$) is not shown for clarity of the plot. The red solid line in the MAGIC panel gives the overall measured flux during the campaign. See text for details.}\label{fig:2344_MWL_LC}
\end{figure*}

\emph{Swift} XRT detected significant variability (see Fig.~\ref{fig:2344_MWL_LC} and Table~\ref{tab:Swift_XRT_results}). The 2\,--\,10\,keV flux increased by $\sim$\,50\,\% within two days, followed by a slow decline nearly halving the flux during eight days. Thereafter, the flux rose again, showing an irregular behaviour, and eventually reached the highest flux during these observations on the last day. The quicklook \emph{Swift} XRT intra-day light curves (from the \emph{Swift} Monitoring Program\footnote{see \href{http://www.swift.psu.edu/monitoring/}{http://www.swift.psu.edu/monitoring/}}) did not show significant intra-day variability during the MW campaign.

Compared to previous observations, also the soft X-ray flux was detected at very low levels during this campaign. In \citet{VERITAS_2344_MWL}, the lowest reported X-ray fluxes from 2\,--\,10\,keV by \emph{Swift} XRT and \emph{RXTE} PCA were $\left(9.6 \pm 0.6\right) \cdot 10^{-12}\,\mathrm{erg}\,\mathrm{cm}^{-2}\,\mathrm{s}^{-1}$ and $\left(9.5 \pm 2.6\right) \cdot 10^{-12}\,\mathrm{erg}\,\mathrm{cm}^{-2}\,\mathrm{s}^{-1}$, respectively. The lowest flux in our sample, which was also used to derive the ``low state'' SED (see Sect.~\ref{chap:SED}), is more than 15\,\% below that level ($\left(7.9 \pm 0.5\right) \cdot 10^{-12}\,\mathrm{erg}\,\mathrm{cm}^{-2}\,\mathrm{s}^{-1}$, see Table~\ref{tab:Swift_XRT_results}). The source is rather often found in such low flux states between 2 and 10\,keV, as further historical measurements show \citep[e.g.\ $8.4 \cdot 10^{-12}\,\mathrm{erg}\,\mathrm{cm}^{-2}\,\mathrm{s}^{-1}$ measured by \emph{BeppoSAX} in 1998, $9 \cdot 10^{-12}\,\mathrm{erg}\,\mathrm{cm}^{-2}\,\mathrm{s}^{-1}$ by \emph{Swift} in 2005;][]{Giommi00,Tramacere07}, but not below the lowest flux reported here.

\begin{figure}
\resizebox{\hsize}{!}{\includegraphics{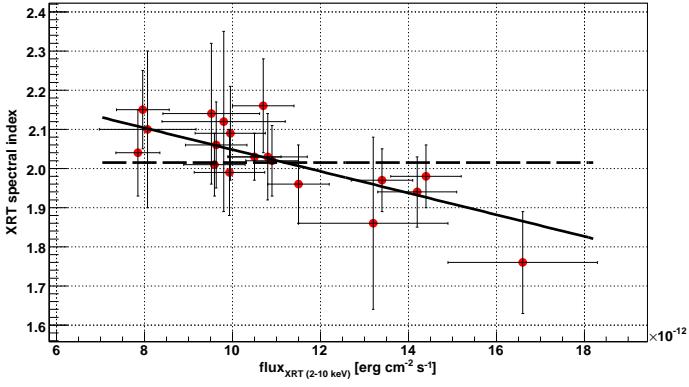}}
\caption{\emph{Swift} XRT spectral index (determined from a simple power law fit between 0.3 and 10\,keV) versus the integral flux for 1ES\,2344$+$514 for this campaign. The dashed line shows a fit with a constant, whereas the solid line denotes a linear fit.}\label{fig:2344_XRT_ind_vs_flux}
\end{figure}

The XRT spectral index (determined from a simple power law fit between 0.3 and 10\,keV) measured during this campaign varied between $1.76 \pm 0.13$ and $2.16 \pm 0.12$, a smaller dynamical range compared to previous observations at similar energies \citep[see e.g.][]{Giommi00,VERITAS_2344_MWL}. Despite not being significantly variable over time ($\chi^2\mathrm{/d.o.f.} = 10.8/18$), there seems to be a dependence of the index on the integral flux, see Fig.~\ref{fig:2344_XRT_ind_vs_flux}, which is produced mainly by the highest measured flux point. A linear fit results in a slope of $-\left(2.77 \pm 1.11\right) \cdot 10^{-2}$ per $10^{-12}\,\mathrm{erg}\,\mathrm{cm}^{-2}\,\mathrm{s}^{-1}$ with a goodness of fit of 99.9\,\% ($\chi^2\mathrm{/d.o.f.} = 4.2/17$), whereas a fit with a constant has a $\chi^2\mathrm{/d.o.f.} = 10.8/18$ (90.5\,\%). A logarithmic likelihood ratio test prefers the linear fit with 97.9\,\%. More meaningful in terms of theoretical models would be to investigate a correlation between the spectral index and the peak position, but because the latter cannot be determined due to lack of significantly curved spectra, the integral flux was used. A negative correlation between flux and spectral index is expected e.g.\ for an increase of the maximum electron energy in SSC models \citep[e.g.][]{Mastichiadis97}.

\begin{figure}
\resizebox{\hsize}{!}{\includegraphics{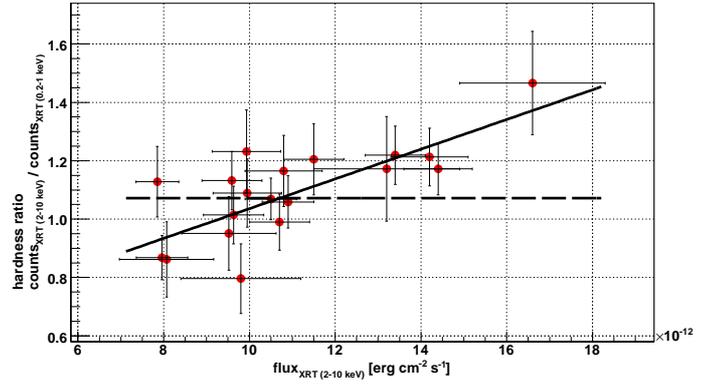}}
\caption{\emph{Swift} XRT hardness ratio versus integral flux for 1ES\,2344$+$514 for this campaign. The dashed line shows a fit with a constant, whereas the solid line denotes a linear fit.}\label{fig:2344_XRT_HR_vs_flux}
\end{figure}

The evolution of the hardness ratio (defined here as the ratio of event counts between 2\,--\,10\,keV and 0.2\,--\,1\,keV), another measure of the spectral shape, cannot be described satisfactorily by a constant fit ($\chi^2\mathrm{/d.o.f.} = 31.6/18$, see Fig.~\ref{fig:2344_MWL_LC}). The detected variability allows to test independently if the spectral shape changed considerably with the flux during the observations. Especially during high flux states, the hardness ratio seemed to increase (judging from Fig.~\ref{fig:2344_MWL_LC}), which means that the flux rose stronger at higher energies than at lower ones. A weak correlation ($\chi^2\mathrm{/d.o.f.} = 12.6/17$ for a linear dependence) between the flux and the hardness ratio is visible (see Fig.~\ref{fig:2344_XRT_HR_vs_flux}). A constant fit yields a $\chi^2\mathrm{/d.o.f.}$ of 31.6/18. Therefore, according to the logarithmic likelihood ratio test, the linear fit is preferred with a confidence of 98.9\,\%. This finding represents an independent confirmation of the correlation between the spectral index and the flux in Fig.~\ref{fig:2344_XRT_ind_vs_flux} and can be interpreted as the common ``harder spectrum when brighter'' trend during a blazar flare \citep[see e.g.][]{HarderWhenBrighter_Pian98}. From earlier observations, 1ES\,2344$+$514 is known to follow such a trend \citep{Giommi00,VERITAS_2344_MWL}.

\begin{figure}
\resizebox{\hsize}{!}{\includegraphics{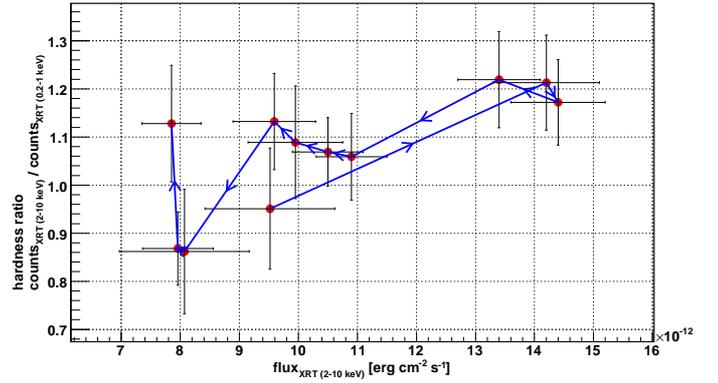}}
\caption{\emph{Swift} XRT hardness ratio versus integral flux for 1ES\,2344$+$514 for the time of the flare (MJD 54757\,--\,54769). The blue arrows give the evolution in time.}\label{fig:2344_XRT_HR_vs_flux_clock}
\end{figure}

Using only the data during the flare, i.e.\ from MJD 54757 to 54769, a hint for a counter-clockwise evolution seems to be apparent in the hardness ratio\,--\,flux plane (Fig.~\ref{fig:2344_XRT_HR_vs_flux_clock}). \citet{Kirk98} explain such a behaviour in a model where the flare arises from a shock front accelerating electrons within a relativistic jet. A counter-clockwise evolution is visible when the observations happen close to the maximum emission frequency of the electrons, where the acceleration and cooling time scales are comparable. In this case, the electrons will not be accelerated to the highest energies and no related flare at gamma-ray energies is expected. This is in agreement with our simultaneous gamma-ray observations, although our VHE light curve does not exclude the presence of a flare of similar amplitude to have appeared at gamma rays at high confidence. An additional hint for similar cooling and acceleration time scales being responsible for the flare is given by the constant spectral index ($\chi^2\mathrm{/d.o.f.}$ of 1.4/10) during the high flux measurements.

Having found a rather hard spectral index down to $\sim$\,1.8 in the XRT band, the BAT 66 months data have been searched for hints of a signal. Indeed, during the time of XRT observations there are indications of a positive flux for several consecutive days, though insignificant due to limited statistics. Therefore, the daily BAT light curve was re-binned to different time scales (see Sect.~\ref{chap:Intro_Swift}). As can be seen from Fig.~\ref{fig:2344_MWL_LC}, variability may be present in the weekly binned data ($\chi^2\mathrm{/d.o.f.} = 30.3/22$ for a constant flux). The weekly high flux point during the XRT measurements at MJD 54772.5 has a significance of 4.9. However, further analysis shows that this anomalous high flux can be attributed to an artifact of the BAT coded-mask imaging and hence is not believed to be due to any real increase in the emission of 1ES\,2344$+$514. For the monthly binned points, the probability of variability increases (constant flux fit: $\chi^2\mathrm{/d.o.f.} = 12.8/4$). The quarterly and yearly results will be discussed in Sect.~\ref{chap:Discussion_interband} in the context of the long-term behaviour of the source.

\subsection{UV and Optical}
KVA found the source on a modest overall flux density level of $\sim$\,4.2\,mJy when compared to earlier and later KVA measurements (see also Fig.~\ref{fig:2344_longterm}). The host galaxy contribution has not been subtracted for the investigation of the light curves. No significant variability throughout the entire observation period is found in the R-band. The data points are consistent with a constant flux density ($\chi^2\mathrm{/d.o.f.} =13.1/29$). The CrAO points are noisier than the KVA points, but all of them are compatible with the KVA data within less than two error bars. Applying a constant fit to the combined KVA$+$CrAO measurements does not provide evidence for variability ($\chi^2\mathrm{/d.o.f.} =35.8/36$). The probability for a constant flux slightly increases for all light curves when subtracting the host galaxy contribution. \emph{Swift} UVOT also did not find significant variability at any of the measured frequencies (see Fig.~\ref{fig:2344_MWL_LC} and Table~\ref{tab:Swift_UVOT_results}).

\subsection{Radio Bands}
\begin{figure}
\resizebox{\hsize}{!}{\includegraphics{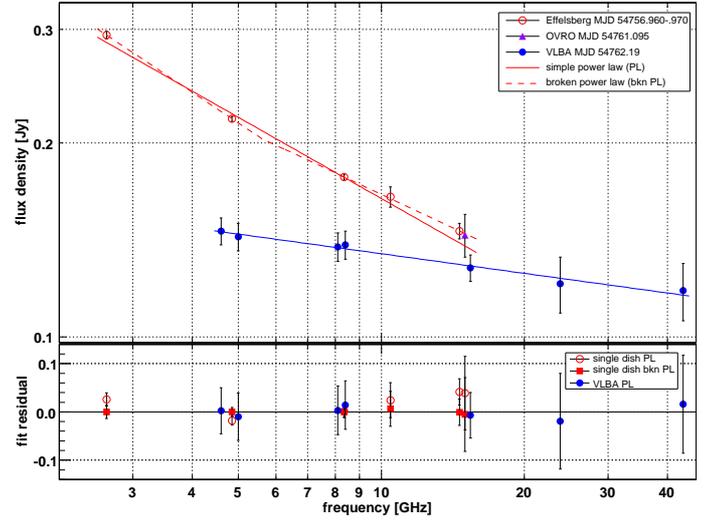}}
\caption{\emph{Top panel:} Radio spectrum of 1ES\,2344$+$514 measured quasi-simultaneously by Effelsberg, OVRO and VLBA. The VLBA points represent integrated flux densities. The solid lines illustrate fits with a simple power law, the dashed line shows a broken power law fit. \emph{Bottom panel:} Fit residuals, which are the differences between the measured flux densities and the fit values divided by the flux density values. See also footnote \ref{ftn:radio}.}\label{fig:2344_radio_VLBA}
\end{figure}

The results of the measurements at radio frequencies have to be discussed in the light of the different observation techniques. The VLBA interferometer is not sensitive to the steep spectrum extended emission from the large scale jet (expected spectral index: $\sim\,0.5$) but observes directly the flat spectrum of the parsec-scale structure, whereas the single-dish telescopes Effelsberg and OVRO measure the whole jet. As the brightness of the extended components decreases at higher frequencies, the parsec-scale spectrum becomes prominent and the single-dish spectrum becomes flatter with increasing frequency. This is obvious from Fig.~\ref{fig:2344_radio_VLBA}, comparing the quasi-simultaneous (separated by $\sim$\,five days) spectra of Effelsberg$+$OVRO and VLBA. Clearly the VLBA integrated spectrum is much flatter than the Effelsberg$+$OVRO spectrum and can be well fitted by a simple power law of the form $S = \nu^{-\alpha}$, where $S$ is the flux density. The resulting spectral index $\alpha$ is $0.10 \pm 0.04$. On the contrary, a simple power law ($\alpha = 0.42 \pm 0.01$) can not describe the Effelsberg$+$OVRO spectrum sufficiently \footnote{Note that all error bars shown in Fig.~\ref{fig:2344_radio_VLBA} contain the systematic contribution, because of which $\chi^2$ goodness of fits cannot be given.\label{ftn:radio}}, judging from the residuals in Fig.~\ref{fig:2344_radio_VLBA}. A broken power law is clearly preferred, whose fit applied to the Effelsberg$+$OVRO data results in the following parameters: $E_\mathrm{break} = \left(5.6 \pm 1.0\right)\,\mathrm{GHz}$, $\alpha_1 = 0.49 \pm 0.03$, $\alpha_2 = 0.34 \pm 0.05$ and a normalisation of $\left(0.153 \pm 0.004\right)$\,Jy at 10\,GHz. $\alpha_1$ being close to 0.5 indicates that the emission is dominated by the large scale jet.

\subsubsection{Single-Dish Observations}\label{chap:Results_radio_single}
Single-dish radio observations were conducted from 2.64\,GHz (Effelsberg) to 228.39\,GHz (IRAM). Since IRAM did not detect the source significantly, 3\,$\sigma$ ULs were calculated. The measurements conducted by Effelsberg show significant variability (although of small amplitude) throughout the observations, as can be seen in the top panel of Fig.~\ref{fig:2344_MWL_LC}. The flux density was rising first towards MJD 54779.0 at all frequencies (observations at 2.64\,GHz were not conducted that day) and declined slowly afterwards. RATAN-600 found the source prior to the Effelsberg observations on a flux density level consistent with the first Effelsberg measurements. The OVRO light curve shows no clear evidence for variability, having a probability of 8.7\,\% ($\chi^2\mathrm{/d.o.f.} = 33.9/24$) for a constant flux density. When omitting the outlier around MJD 54870, the probability for a constant flux density is rising to 13.6\,\%. 1ES\,2344$+$514 was too faint to be detected by Mets\"ahovi during the campaign and for 07/10/2008 (MJD 54746), an upper limit on the flux density at 37\,GHz of $< 0.33$\,Jy with S/N $> 4$ was calculated. The source was detected by Mets\"ahovi three months earlier at a flux density level of $\left(0.38 \pm 0.09\right)$\,Jy, which is consistent with the derived upper limit.

\begin{figure}
\resizebox{\hsize}{!}{\includegraphics{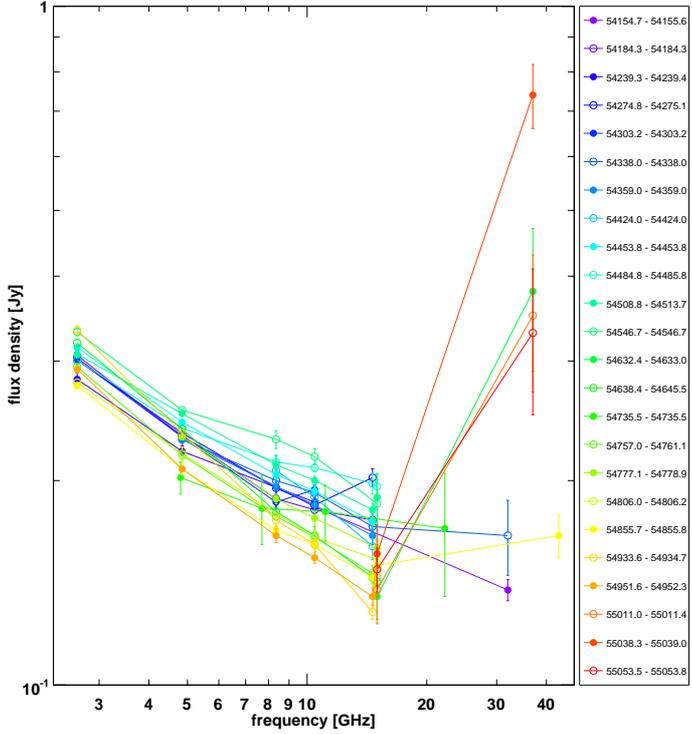}}
\caption{Radio spectra of 1ES\,2344$+$514 taken by Effelsberg, Mets\"ahovi, OVRO and RATAN-600 from 2007 through 2009. Data from different instruments have been combined if the time difference was less than 14\,days. The thin solid lines simply connect the data points. The legend contains the MJD range of the combined spectra. We recall that the detection limit of Mets\"ahovi at 37\,GHz under optimal conditions is $\sim$\,0.2\,Jy. See also Fig.~\ref{fig:2344_radio_spectra_single}.}\label{fig:2344_radio_spectra}
\end{figure}

To understand the radio behaviour of AGNs, they have to be studied over long periods of time, considering the rather long variability time scales compared to e.g.\ X-rays. 1ES\,2344$+$514 has been observed in the past on a regular basis at radio frequencies. The combined quasi-simultaneous (time difference $< 14$\,days) radio spectra from Effelsberg, Mets\"ahovi, OVRO and RATAN-600 from 2007 through 2009 are shown in Fig.~\ref{fig:2344_radio_spectra} (for a time-resolved version see Fig.~\ref{fig:2344_radio_spectra_single}). IRAM ULs, where the lowest flux density UL is 0.96\,Jy, are not shown for clarity. At frequencies below $\sim$\,20\,GHz the source shows steep radio spectra while above this frequency, the spectra become flat or inverted. This is a consequence of high amplitude variability of the mm radio emission, originating from a more compact region than the one dominating the cm-band radio spectrum. These characteristics are in accordance with the model of \citet{Angelakis12} who demonstrated that the radio spectra of most of the AGNs under study can be described well by a simple two component system consisting of a power-law quiescent spectrum (attributed to e.g.\ the optically thin diffuse emission of a large scale jet) and a convex synchrotron self-absorbed spectrum (resulting from a recent outburst within the compact region).

Such an outburst may be explained in the framework of the model of \citet{Marscher85} where the emission is coming from a shock propagating in an adiabatic relativistic jet. According to shock models, the feature should move outwards within the jet, i.e.\ from high to low frequencies. Outbursts are present only at times outside of the principal MW campaign, the most significant one seen by Mets\"ahovi around MJD 55039 having a doubling time of $\lesssim 28$\,days and a decline to the original flux density value of $\lesssim 15$\,days. IRAM observations two days later provided only unconstraining flux density ULs ($< 1.74$\,Jy at 86.24\,GHz and $< 1.95$\,Jy at 142.33\,GHz), and the quasi-contemporaneous OVRO points, from the flaring day as well as 4, 12 or 15\,days after the flare, did not show a significantly higher flux density. However, the flare may have been missed due to the comparably sparse sampling during these days. A fit with a constant to the OVRO data from MJD 55024 until MJD 55054 does not yield significant variability ($\chi^2\mathrm{/d.o.f.} = 4.9/8$). Hence no conclusions on the validity of the shock scenario can be drawn from this data. However, the time scale of the flare itself is interesting. There are very few examples of such fast variability at 37\,GHz for HBLs, e.g.\ Mrk\,421 \citep{Lichti08}. That is mainly due to their faintness and consequently low detection rate at this frequency. Nevertheless some of these objects are detected at clearly higher flux density values in between periods of non-detections, giving a hint for fast variability \citep[see e.g.][]{Nieppola07}.

\begin{figure}
\resizebox{\hsize}{!}{\includegraphics{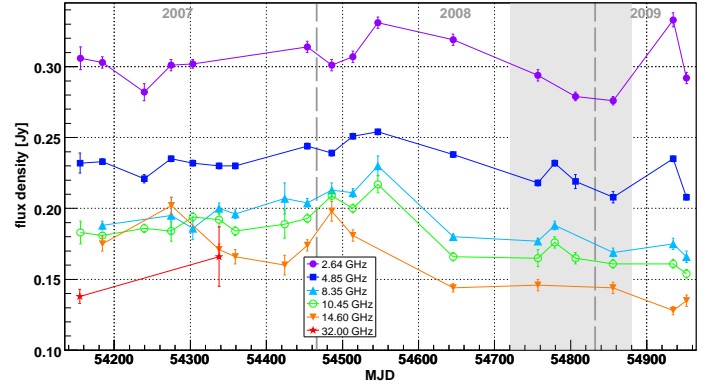}}
\caption{Light curve of 1ES\,2344$+$514 measured by Effelsberg. The solid lines connect the points for illustrative purposes. The MW campaign duration is visualised by the grey box.}\label{fig:2344_Effelsberg_lightcurve}
\end{figure}

Figure~\ref{fig:2344_Effelsberg_lightcurve} shows the light curve measured by Effelsberg in the context of the F-GAMMA program from beginning of 2007 until mid of 2009 (MJD 54155\,--\,54952). Apart from an overall higher flux density state especially at high frequencies from the start of the observations until mid of 2008 ($\sim$\,MJD 54600), several structures are visible, which reveal an ambivalent correlation during the variability. On the one hand, there are features where only near-by frequency bands showed the same trends with gradually decreasing tendency (MJDs 54239, 54486 and $\sim$\,54935). On the other hand, structures may have been present at all measured bands with about the same strength (MJDs $\sim$\,54547 and $\sim$\,54779). The different features are possibly attributed to re-acceleration of particles within the jet. The occurrence of an equal amplitude at all frequencies or gradually decreasing amplitude with frequency can be explained in that context by different physical conditions within the jet, e.g.\ a change of the magnetic field or the particle density. Alternatively, the sparse sampling in combination with frequency-dependent time lags may explain some of the observed features.

The probability that the flux density seen by OVRO was constant during the second Effelsberg high state, between MJD 54761 and 54796, is rather low ($\chi^2\mathrm{/d.o.f.} = 21.2/12$, i.e.\ $\sim$\,4.8\,\%), due to the first measurement in this time period (see also Fig.~\ref{fig:2344_MWL_LC}). Neglecting this point, the $\chi^2\mathrm{/d.o.f.} = 2.8/11$, giving highly significant evidence for constancy. The flux density rose from the first point within two days by $\sim$\,16\% and remained constant afterwards. This indicates that the peak seen in the Effelsberg light curve around MJD 54779 was indeed a broad high flux density plateau. From the OVRO variability time scale, the Doppler factor can be estimated to be $> 3.4$ using Eqs.~(1) and (2) in \citet{Doppler}. It should be noted that this estimation method has not been tested for faint radio sources like TeV BL Lacs, and that the estimation of the flare rise time is based on two data points only. Therefore, the value is not representative, but also not in disagreement with the quasi-simultaneous results from the high state SED modelling (see Sect.~\ref{chap:SED}).

\subsubsection{Interferometric Observations}
\begin{figure}
\resizebox{\hsize}{!}{\includegraphics[angle=-90]{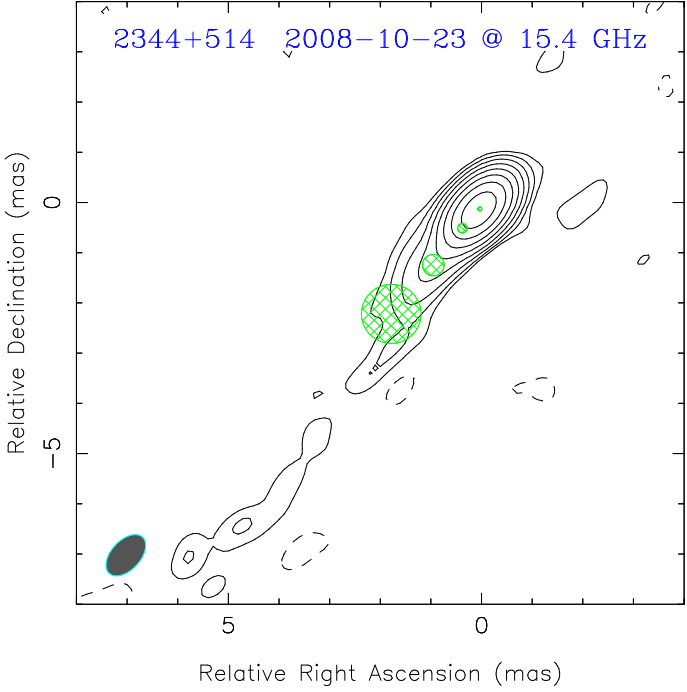}}
\caption{Naturally weighted CLEAN image of 1ES\,2344$+$514 obtained with the VLBA on 23/10/2008 (MJD 54762) at 15.4\,GHz. Green circles mark positions and sizes of model components listed in the Table~\ref{tab:jet_components} from North-West to South-East: the Core, C\,3, C\,2, C\,1. 1\,mas corresponds to 0.9\,pc in absolute length. Contour image parameters: $\mathrm{peak} = 0.11$\,Jy/beam, $\mathrm{beam} = 0.98 \times 0.58$\,mas at $\mathrm{PA} = -42.2^\circ$, first $\mathrm{contour} = 0.50$\,mJy/beam, contour level increase $\mathrm{factor} = 2$. The naturally weighted beam is shown at the lower left corner of the image.}\label{fig:2344_MOJAVE}
\end{figure}

The VLBA image of 1ES\,2344$+$514 (Fig.~\ref{fig:2344_MOJAVE}) reveals a core-dominated structure with a smooth jet extending in South-East direction. At the distance of 1ES\,2344$+$514 ($z = 0.044$), the linear scale of the images is 0.9\,pc/mas. The integrated parsec-scale radio spectrum is flat, which is typical for blazars. The VLBA data can be used to estimate the radio core size (the compact feature at the North-Western end of the jet in Fig.~\ref{fig:2344_MOJAVE}) at each frequency (see Table~\ref{tab:coresize}) by modelling it with a circular Gaussian emission component. At the highest and lowest frequency the core size can only be constrained to $< 0.13\,\mathrm{mas}$, i.e.\ $R \lesssim 10^{17}$\,cm, while at the other frequencies the core is resolved. The limiting resolution and component size uncertainties were estimated following \citet{Fomalont99}, \citet{Lobanov05} and \citet{Kovalev05}.

\begin{table}
\caption{Core size as a function of frequency $\nu$. Errors are given with 1\,$\sigma$.}\label{tab:coresize}
\centering
\begin{tabular}{llll}
\hline\hline
$\nu$ & Flux Density & Size & Resolution Limit\\
{[}GHz] & [Jy] & [mas] & [mas]\\\hline
\phantom{0}4.6  & $0.094 \pm 0.006$ & $<0.13$ & 0.13\\
\phantom{0}5.0  & $0.091 \pm 0.005$ & $0.16 \pm 0.09$ & 0.10\\
\phantom{0}8.1  & $0.094 \pm 0.005$ & $0.09 \pm 0.06$ & 0.06\\
\phantom{0}8.4  & $0.096 \pm 0.005$ & $0.14 \pm 0.05$ & 0.06\\
15.4  & $0.102 \pm 0.007$ & $0.07 \pm 0.04$ & 0.05\\
23.8  & $0.076 \pm 0.005$ & $0.03 \pm 0.02$ & 0.03\\
43.2  & $0.120 \pm 0.022$ & $<0.06$ & 0.06\\\hline
\end{tabular}
\end{table}

The source is highly core dominated at parsec scales. Specifically, on the basis of modelling of VLBI data we can estimate that the emission from the core region at 5\,GHz accounts for $\sim$\,65\,\% of the total VLBI flux density progressively increasing up to $\sim$\,75\,\% at 23.8\,GHz. Fast variations with rms values typically well below 10\,\% in total flux density have been measured at cm wavelengths in a large sample of flat-spectrum compact radio sources \citep[e.g.,][]{Kraus03,Lovell08} -- most probably due to scintillation in the Galactic interstellar medium.

The flat parsec-scale radio spectrum showing no clear signs of the synchrotron self-absorption turnover at low frequencies (see Fig.~\ref{fig:2344_radio_VLBA}) may be explained as optically thin synchrotron emission from an ensemble of electrons having a very hard energy spectrum $N(E) \propto E^{-X}$ \citep{Sokolovsky_magnetic_field}. However, the more likely explanation is that the flat spectrum is a result of optically thick synchrotron emission from a \citet{BlandfordKoenigl79} type jet. This explanation is supported by the observed core size increase at lower frequencies \citep[][see Table~\ref{tab:coresize}]{Unwin94} and the difference in separation between the component C\,3 and the core observed at 15.4 \citep{Piner04} and 43.2\,GHz \citep{Piner10}. Together, these points agree with the standard interpretation of the parsec-scale radio core in 1ES\,2344$+$514 as a surface in a continuous \citet{BlandfordKoenigl79} jet at which the optical depth at a given frequency $\nu$ is $\tau_\nu \simeq 1$ \citep{Lobanov98,Sokolovsky_synchrotron}. This is a challenge to most of the alternative interpretations of the core physics discussed by \citet{Marscher06,Marscher08}, at least for the frequency range covered by our VLBA observations.

Using multi-epoch MOJAVE results (see Table~\ref{tab:jet_components}), the average core brightness temperature at 15.4\,GHz can be determined as $T_\mathrm{b} \approx 8 \cdot 10^{10}$\,K. While being rather smooth, the jet of 1ES\,2344$+$514 can still be divided into several individual emission components that we fit with circular or elliptical Gaussian models. Consistency of their positions, fluxes and sizes among MOJAVE epochs suggests that these components are real stable structures in the jet, not an artefact of representation of a smooth continuous jet with discrete Gaussian components. Analysis of the 15.4\,GHz MOJAVE monitoring shows no significant motion of the jet components over the entire observing period of eleven years. Even across the long eight-year time gap, the positional changes of the fitted component positions are smaller than their overall scatter in the post-2008 period. Parameters of the jet components and results of the formal linear fits to their trajectories are presented in Table~\ref{tab:jet_components}. Among the jet components, C\,3 is the brightest and smallest one, located $\sim$\,0.6\,mas from the core. C\,3 provides the strongest limits on the apparent jet speed $v_\mathrm{app}$ of $\left(-5 \pm 7\right)\,\mu\mathrm{as\,yr^{-1}}$ corresponding to $\beta_\mathrm{app} = v_\mathrm{app}/c = -\left(0.01 \pm 0.02\right)$. It can be clearly identified with the component C\,3 described in \citet{Piner04} and \citet{Piner10}, where the $\beta_\mathrm{app}$ values derived for C\,3 were given as $-0.19 \pm 0.40$ and $0.10 \pm 0.02$, respectively.

That no superluminal motion is observed in the jet of 1ES\,2344$+$514 is in line with the previous findings that this source and a number of other TeV \citep{Piner04,Piner10} and non-TeV \citep{Karouzos12} BL Lacs show much slower apparent jet speeds compared to those typically found in compact extragalactic radio sources \citep{Lister09_VI}. Note however, that \citet{Piner10} report the detection of significant component motion in 1ES\,2344$+$514 with speeds inconsistent with the results presented above. The possible sources of this discrepancy include (i) the smaller number of observational epochs available to \citet{Piner10} and (ii) the fact that the authors combine component positions measured at different frequencies without explicitly taking the effect of a frequency-dependent core shift into account \citep{Lobanov98,Sokolovsky_synchrotron,Hada11}, which may introduce systematic errors in the positions of the components.

There is a possibility that the observed jet component motion is not indicative of the actual jet flow speed in this source. However, that assumption is not supported by the fact that the core brightness temperature is well below the inverse Compton limit $\sim$\,10$^{12}$\,K \citep{Kellermann69,Kovalev05}. The rather low brightness temperature of the core indicates that the radio emitting plasma in the jet is probably affected by only moderate relativistic beaming.

\section{Discussion}\label{discussion}
\subsection{Cross-Band Correlations and Variability Studies}\label{chap:Discussion_interband}
\begin{figure}
\resizebox{\hsize}{!}{\includegraphics{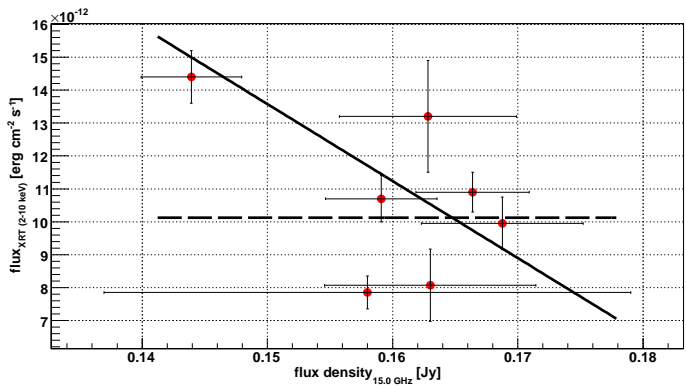}}
\caption{Integral flux measured by \emph{Swift} XRT from 2\,--\,10\,keV versus the flux density measured by OVRO for data pairs with a time difference of $< 0.9$\,days. A constant fit is represented by the dashed line, the solid line shows the result of a linear fit.}\label{fig:2344_OVRO_XRT}
\end{figure}

Correlated variability at different energy bands or lack of such provides important information on the emission mechanisms and locations. During the MW campaign, significant variability was only present in the low frequency radio and X-ray regime. The small and short flux density rise at the beginning of the OVRO measurements was accompanied by an XRT flux declining from the flare peak down to $\sim$\,75\,\%, suggesting an anti-correlation between these two frequency bands. Mathematically, a linear dependence (probability of 45.2\,\%, $\chi^2\mathrm{/d.o.f.} = 4.7/5$) is clearly preferred above a constant one (logarithmic likelihood ratio test probability in favour of the linear fit is 98.9\,\%), see Fig.~\ref{fig:2344_OVRO_XRT}. Any possible correlation is, however, dominated by the point with the lowest flux density, hence the OVRO\,--\,XRT comparison is inconclusive.

Since the MAGIC light curve points are only marginally significant, the feasibility of investigating correlations with other bands is limited. It should however be noted that the daily flux point with the highest significance in the MAGIC light curve appeared only $\sim$\,two days after the highest \emph{Swift} XRT flux had been detected. Since this high X-ray flux was accompanied by a strong hardening of the spectral index, this can be interpreted as an injection of fresh electrons into the emission region, which should cause a higher flux also at gamma-ray energies.

\begin{sidewaysfigure*}
\centering
\rule{0.\textheight}{0.745\textheight}
\resizebox{\hsize}{!}{\includegraphics{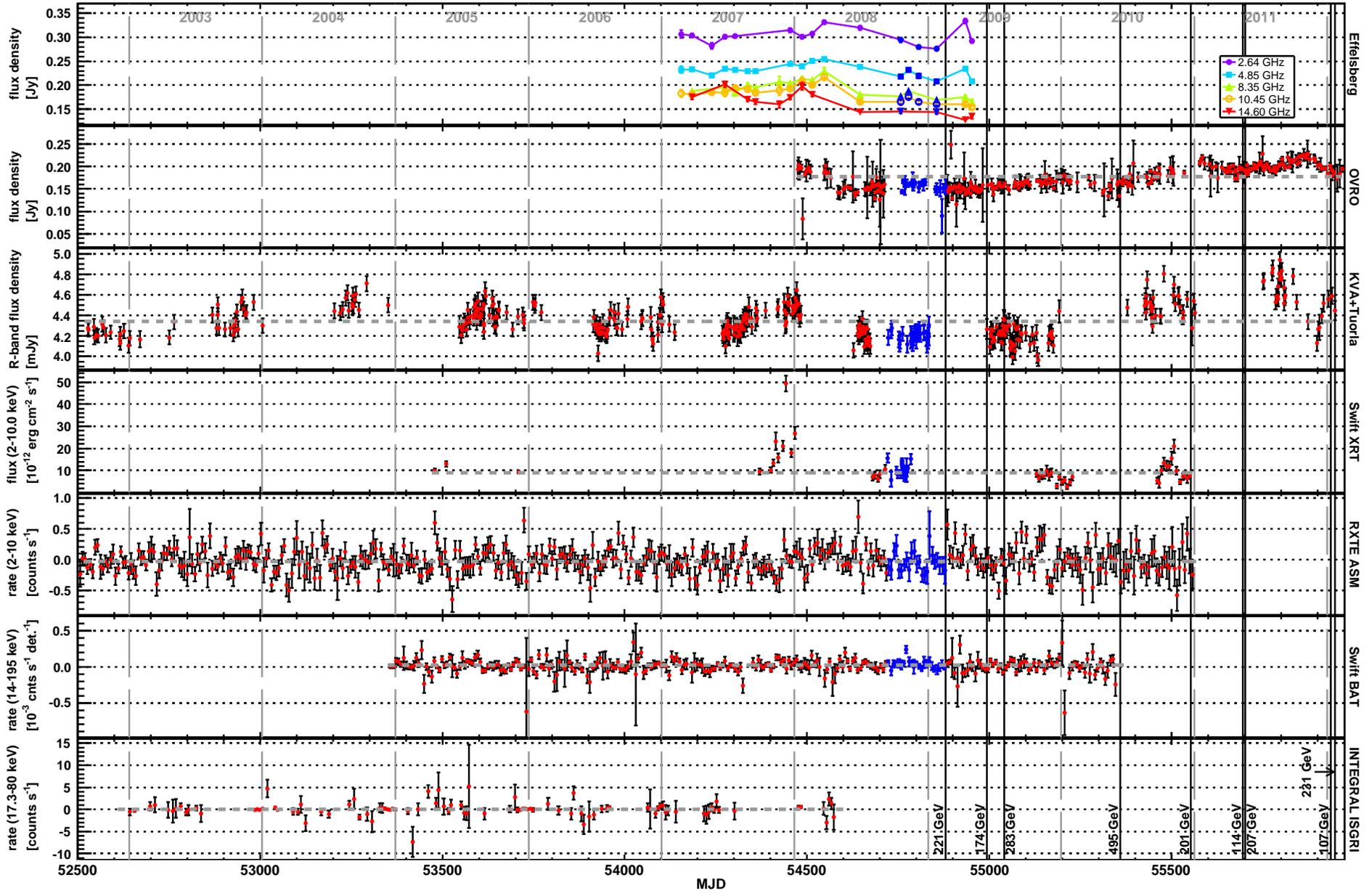}}
\caption{Long term light curves of 1ES\,2344$+$514 measured by Effelsberg, OVRO, KVA$+$Tuorla, \emph{Swift} XRT, \emph{RXTE} ASM, \emph{Swift} BAT and \emph{INTEGRAL} ISGRI. The first four panels show an observation-wise binning ($< 1$\,day), the last three panels have been re-binned to weekly scales. The blue points mark the MJD range shown in Fig.~\ref{fig:2344_MWL_LC}. The lines in the top panel do not result from a fit but simply connect the data points. Horizontal short-dashed lines give fits with a constant. The black solid vertical lines denote the arrival time of LAT events $> 100$\,GeV. The ASM light curve has been cut at 01/01/2011. For clarity, two points of the BAT light curve are not shown (MJD 54177.5, rate $\left(2.26 \pm 0.91\right) \cdot 10^{-3}\,\mathrm{cnts}\,\mathrm{s}^{-1}$ per detector; MJD 54541.5, rate $\left(3.16 \pm 1.82\right) \cdot 10^{-3}\,\mathrm{cnts}\,\mathrm{s}^{-1}$ per detector) and the (linear) error bar of one point (MJD 53729.5) has been cut at the plot edge. Distribution histograms of the light curve points over their respective errors are given in Fig.~\ref{fig:2344_longterm_hist}. See text for details.}\label{fig:2344_longterm}
\end{sidewaysfigure*}

Considering time scales beyond the duration of this MW campaign, flux (density) changes in the radio, optical and X-ray regime were clearly detected for 1ES\,2344$+$514, as can be seen from the first four light curves in Fig.~\ref{fig:2344_longterm}. A fit with a constant results in $\chi^2\mathrm{/d.o.f.} = 3929.6/317$ for OVRO, $\chi^2\mathrm{/d.o.f.} = 2058.3/437$ for KVA and $\chi^2\mathrm{/d.o.f.} = 491.8/66$ for \emph{Swift} XRT. Also the distribution histograms of flux (density) over error (Fig.~\ref{fig:2344_longterm_hist}) show a clear deviation from a Gaussian function, where the latter would be expected for uniformly sampled light curves dominated by statistical fluctuations. The strong flare measured by XRT around MJD 54442.2 cannot be unambiguously identified in the Effelsberg or KVA light curves. In the KVA light curve, a slightly higher flux was seen $\sim$\,5.3 days after the large flare and $\sim$\,5.7 days after a smaller XRT flare (MJD 54466.5), suggestive of a time lag of the optical emission with respect to the X-rays. Also the Effelsberg measurements revealed two significant peaks (at different frequencies, though) in that time period ($\sim$\,MJDs 54486 and 54547). But on a significant correlation of these high states with the XRT flares can only be speculated due to the incomplete sampling of the \emph{Swift} XRT, KVA and Effelsberg light curves.

On time scales of years, the good sampling allows us to perform a search for correlations between the OVRO and KVA data. In order to exclude a bias of the result caused by measurement noise, OVRO data with an error $> 0.02$\,Jy ($\sim$\,15\,\% of the data) were excluded from this analysis. Using the discrete correlation function (DCF) as defined in \citet{Edelson88}, we searched for possible correlations for lags up to $\pm 100$\,days between both data sets. Two such searches were performed, one in which the raw light curves were used, and one in which we searched for correlations after first subtracting off a low-pass filtered version of the data in order to remove long-term trends which might influence the calculation of the DCF. The analysis did not yield a significant correlation.

Investigating the light curves of \emph{RXTE} ASM\footnote{\emph{RXTE} ASM data were obtained from NASA GSFC's archive. In the generation of the light curve only single-dwell ASM data were used in which the $\chi^2_{red}$ was $< 1.3$. Slight variations in the signal to noise ratio over the full ASM light curve are due to episodes where the source position was less well covered by the individual cameras of the ASM. Due to a strongly reduced instrument sensitivity resulting from degradation of the detectors towards the end of the mission, no data since 01/01/2011 have been used.}, \emph{Swift} BAT and \emph{INTEGRAL} ISGRI\footnote{Data taken from HEAVENS \citep[][\href{http://www.isdc.unige.ch/heavens/}{http://www.isdc.unige.ch/heavens/}]{HEAVENS}.} on time scales of one day, some outliers become apparent. These are expected from a statistical point of view, and all data points but one do not have a signal/error significantly offset from their corresponding Gaussian distribution (see Fig.~\ref{fig:2344_longterm_hist_daily}). This point, having a signal to noise ratio of $\sim$\,5, was measured by ASM at MJD 54468.0, 1.5\,days after an increased flux seen by \emph{Swift} XRT and $\sim$\,3.8\,days before the higher KVA state (see above). If real, it indicates that XRT detected the onset of a flare potentially even higher than the large one around MJD 54442.2, which preceded a small flare in the R-band by $\sim$\,3\,--\,4\,days. The sparse sampling does not allow to draw further conclusions on the nature or origin of the flare.

A fit with a constant to the daily light curves is ruled out on high statistical basis for ASM and BAT ($\chi^2\mathrm{/d.o.f.} = 4215.6/3007$ and 2364.1/1733, respectively), though not for ISGRI ($\chi^2\mathrm{/d.o.f.} = 228.8/214$). The Gaussian fits to the signal/error distributions reveal a significant shift of the mean value to positive values for BAT and ISGRI, but not for ASM ($0.18 \pm 0.03$, $0.11 \pm 0.07$ and $-\left(0.07 \pm 0.02\right)$ for BAT, ISGRI and ASM, respectively). Note, however, that a Gaussian statistical behaviour is not expected for ASM due to coded mask observations. In the case of BAT, Gaussian statistics is still applicable despite applying the coded mask technique due to the large number of individual detector elements. Consequently, the BAT light curve indicates significant variability of the source at hard X-rays.

The large flare detected by XRT on MJD 54442.2 is not clearly visible in the daily ASM or BAT light curve; ISGRI did not observe at that time. 1ES\,2344$+$514 seems to be too faint at X-rays to be detected on daily time scales by these two instruments. Therefore, the light curves of ASM, BAT and ISGRI have been re-binned in the same way as the simultaneous BAT data (see Sect.~\ref{chap:Intro_Swift}). As an example, the weekly results are shown in Figs.~\ref{fig:2344_longterm} and \ref{fig:2344_longterm_hist}. From the signal/error distribution histograms, no significant flares outside the Gaussian distribution\footnote{A closer look at the light curves reveals several extended peaks with rather symmetrical rise and fall times in the ASM and BAT data for different binnings, but these structures match quite well the minima of the solar angle to ASM and can therefore most probably not be ascribed to 1ES\,2344$+$514.} are apparent for any binning and instrument except for the 1-week BAT point at MJD 54772.5 already discussed in Sect.~\ref{chap:Results_Swift}. The large XRT flare at MJD 54442.2 is still not clearly present in any of the light curves for any binning.

The trend of a positive Gaussian mean increases with larger time bins for BAT, finally leading to the detection of 1ES\,2344$+$514 as reported in the 58-Month Catalog. The light curves of ASM and BAT are not consistent with a constant flux up to quarterly binning, though the corresponding probabilities are rising with increasing time bin size. For a yearly binning, a fit with a constant yields $\chi^2\mathrm{/d.o.f.} = 14.9/7$, $\chi^2\mathrm{/d.o.f.} = 5.1/5$ and $\chi^2\mathrm{/d.o.f.} = 8.2/5$ for ASM, BAT and ISGRI, respectively.

The arrival times of the nine photons with energy $> 100$\,GeV detected by \emph{Fermi}-LAT within the first 44 months are also shown in Fig.~\ref{fig:2344_longterm}. No exceptional behaviour is visible in the light curves of the other energy bands at these times.

At the time of the Mets\"ahovi flare (MJD 55039, see Sect.~\ref{chap:Results_radio_single}), optical R-band monitoring data are also available. 1ES\,2344$+$514 was densely covered between MJDs 54994 and 55088 with detections on daily basis from seven days before the Mets\"ahovi flare until three days after. The best, though still insignificant, hint of a higher flux density in that period is given at MJDs 55040 and 55041. Either there was no correlation present between the R-band and 37\,GHz during this flare (as the missing long-term OVRO\,--\,KVA correlation is also suggesting), or the optical monitoring missed it. A missing correlation would hint on different flaring mechanisms or, more likely, spatially separated emission regions. In the latter case, a time delay between the radio and optical emission would be expected, which may be more firmly determined based on continuous monitoring in the future. Quite interestingly, two of the \emph{Fermi}-LAT events with energies $> 100$\,GeV were detected 46 days before and $< 2$\,days after the Mets\"ahovi flare, respectively (see Table~\ref{tab:FermiHE}). On a correlation can only be speculated though, taking into account that the exact time of the flare maximum can be determined neither from the Mets\"ahovi nor the OVRO or KVA monitoring data. Moreover, the timing of the events may be purely coincidental.

In general, the present monitoring programs at various wavelengths represent a major progress towards the understanding of blazar phenomena. Nevertheless, more efforts seem necessary, increasing the sampling density and time basis, and especially extending the monitored energy range to the X-ray and VHE regime.

\subsection{Spectral Energy Distribution Modelling}\label{chap:SED}
\subsubsection{Simultaneity}
Since the gamma-ray detections were only marginally significant, (quasi-)simultaneous data sets for constructing SEDs are composed according to the X-ray flux state. We define a ``low'' and ``high'' X-ray flux SED, choosing MJDs 54760.9 (high) and 54768.8 (low). The exact observation times of the different instruments around these data sets are given in Table~\ref{tab:sim_observations}.

\begin{table}
\caption{Observing intervals of the SED data sets.}\label{tab:sim_observations}
\centering
\begin{tabular}{lll}
\hline\hline
Data Set & Instrument & Observation Time\\
& & [MJD]\\\hline
\multirow{6}*{1ES\,2344$+$514 low} & Effelsberg & 54778.947\,--\,54778.950\\
& OVRO & 54769.077\\
& \emph{Swift} & 54768.806\,--\,54768.948\\
& AGILE & 54770.500\,--\,54800.500\tablefootmark{a}\\
& \emph{Fermi}-LAT & 54759.941\,--\,54800.897\tablefootmark{a}\\
& MAGIC & 54759.941\,--\,54800.897\\\hline
\multirow{8}*{1ES\,2344$+$514 high} & Effelsberg & 54756.960\,--\,54756.970\\
& OVRO & 54761.095\\
& VLBA & 54761.96\phantom{0}\,--\,54762.42\\
& KVA & 54761.718\\
& \emph{Swift} & 54760.899\,--\,54760.983\\
& AGILE & 54770.500\,--\,54800.500\tablefootmark{a}\\
& \emph{Fermi}-LAT &  54759.941\,--\,54800.897\tablefootmark{a}\\
& MAGIC & 54759.941\,--\,54800.897\\\hline
\end{tabular}
\tablefoot{
\tablefoottext{a}{No detection; 95 \% c.l.\ ULs calculated.}
}
\end{table}

The two data sets are too close in time to derive individual gamma-ray results. The corresponding ULs of AGILE-GRID and \emph{Fermi}-LAT as well as the MAGIC spectrum were averaged over the entire respective observation periods and used for the modelling of both SEDs. Note that no significant variability could be found in any gamma-ray band (see Sects.~\ref{chap:Results_VHE},\ref{chap:Results_HE}), the detection of which would exclude averaging the measurements.

The 66-month BAT spectrum can be regarded as a measure of the average low flux of the source, since no significant flares are present in the BAT light curve from daily to yearly scales apart from the artificially high weekly flux point at MJD 54772.5. Taking into account that the low and high state XRT spectra have been measured in a slightly lower and higher flux state than the long-term average (judging from Fig.~\ref{fig:2344_longterm}), respectively, the BAT spectrum may be considered as being quasi-simultaneous to these spectra.

The variability time scale during the observations in the radio regime is hard to assess and differs from band to band, but in general large changes in flux are not expected on time scales of $\sim$\,two weeks. Therefore, some radio measurements have been included on a quasi-simultaneous basis.

For the chosen low and high state dates of 1ES\,2344$+$514, there were no simultaneous measurements by KVA. For the high state, the result from the following night is used. Taking into account that the KVA measurements do not show a hint of variability throughout more than three months of observations, this procedure seems justified. On the other hand it should be noted that the simultaneous optical\,--\,X-ray data pairs are very few and especially missing for high X-ray fluxes, hence a higher optical flux during the XRT flare cannot be excluded.

\subsubsection{Model Description}
Two different leptonic SSC emission models have been applied to the such defined quasi-simultaneous SEDs. The one-zone model by \citet{MaraschiTavecchio03} describes the SED completely by nine parameters: the radius $R$, Doppler factor $\delta$ and magnetic field $B$ of the emission region, which contains an electron distribution following a broken power law with index $n_1$ for $\gamma_\mathrm{min} < \gamma < \gamma_\mathrm{break}$ and index $n_2$ for $\gamma_\mathrm{break} < \gamma < \gamma_\mathrm{max}$ with density $K$ at Lorentz factor $\gamma = 1$. $\gamma_\mathrm{min}$ has been fixed to values of 1 and 4000, which represent the extreme cases of the lowest and a very high realisation, visualising a large part of the reasonable parameter range.

In the second model \citep{Weidinger10}, an electron distribution with density $K$ at $\gamma_\mathrm{min}$ is being accelerated in a zone with radius $R_\mathrm{acc}$. The electrons are finally escaping the acceleration region and enter a second region, i.e.\ the emission region, with radius $R_\mathrm{em}$ where no further acceleration takes place. The magnetic field $B$ and Doppler factor $\delta$ are the same for the two regions. The resulting electron distribution as well as the spectral indices are derived self-consistently from the acceleration and cooling processes and are not determined a priori.

\subsubsection{Results}\label{chap:SED_results}
\begin{figure*}
\resizebox{\hsize}{!}{\includegraphics{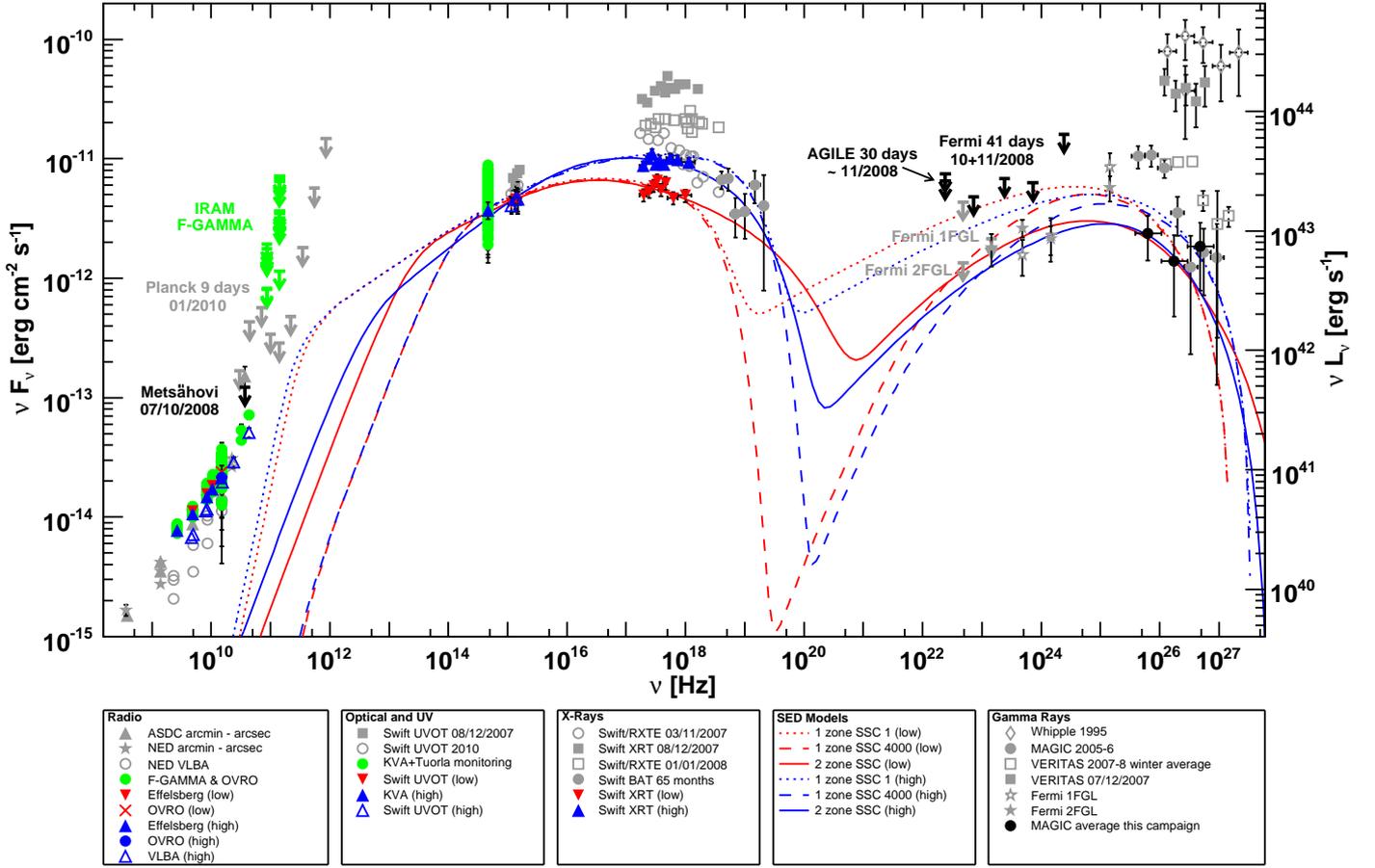}}
\caption{Simultaneous SEDs of 1ES\,2344$+$514 as derived from this campaign (black, blue and red markers) together with monitoring results from F-GAMMA and the Tuorla Blazar Monitoring Program (green points) as well as archival data (shown in grey). The black MAGIC data points represent the overall flux during the campaign. ``low'' and ``high'' denote the simultaneous data sets as given in Table~\ref{tab:sim_observations}. For the one-zone model, the number given in the legend represents the value of $\gamma_\mathrm{min}$. See text for further details. Archival data have been taken from the ASDC SED Builder (\href{http://tools.asdc.asi.it}{http://tools.asdc.asi.it}), \citet{VERITAS_2344_MWL}, \citet{Whipple_2344} (Whipple) and \citet{MAGIC_2344} (MAGIC). The \emph{Planck} ULs were taken from \citet{Giommi11}.}\label{fig:2344_SED}
\end{figure*}

The 1ES\,2344$+$514 SEDs compiled from this campaign are shown in Fig.~\ref{fig:2344_SED}. The given simultaneous KVA and UVOT data have been host-galaxy corrected where necessary as well as de-reddened. Due to the strong host galaxy and the large uncertainty in its flux, the V-, B- and U-band fluxes cannot be determined with sufficient significance and hence are not shown. Since the optical data given at ASDC are not host-galaxy corrected, they have been omitted. The simultaneous XRT data were corrected for Galactic absorption. MAGIC and Whipple data were EBL de-absorbed using the model from \citet{Kneiskelow}, the archival VERITAS data by the \citet{EBL_Franceschini08} model. The AGILE UL denotes the flux $> 100$\,MeV, the \emph{Fermi}-LAT 1FGL ULs are given between 0.1 and 0.3 as well as 0.3 and 1\,GeV. Note that the Mets\"ahovi UL has a c.l.\ of 4\,$\sigma$ and the IRAM ones 3\,$\sigma$.

A comparison of our results to archival SED data reveals that the source has been measured in one of the lowest flux states ever obtained from X-ray to VHE gamma rays. At optical and radio frequencies, the fluxes were at a modest level.

For comparison, we also include published \emph{Fermi}-LAT 1FGL points on a quasi-simultaneous basis, taking into account that no significant variability is present also in 2FGL. Though all points are consistent with a simple power law distribution, there seems to have been a small jump between the LAT points at 6 and 60\,GeV which is hard to describe with the applied models. Either the 6\,GeV point is rather low in flux or the 60\,GeV point comparably high. The latter point connects rather smoothly to the MAGIC spectrum from 2005\,--\,6 and the VERITAS points from 2008, whereas putting more weight on the 6\,GeV point the MAGIC spectrum from this campaign seems to match better the quasi-simultaneous LAT data. Since the highest energy point is subject to a rather large statistical uncertainty ($< 10$ and $< 20$\,events in 1FGL and 2FGL, respectively) and hence prone to potential short-term flux variations, our modelling is focusing on the 0.6\,--\,6\,GeV points. In this context, the inconsistency of the \emph{Fermi}-LAT spectrum with the VHE points reported in \citet{Fermi_TeVselected} is no longer evident using the MAGIC points from this campaign, indicating that the VHE results derived here are more representative of the average flux state of 1ES\,2344$+$514.

The 66-month BAT spectrum is an adequate extrapolation of the XRT high state from this campaign. The low state spectrum, on the other hand, would require an increase of flux with rising energy, which cannot be described with the SSC models applied here. Therefore, the BAT spectrum has been considered for the fit to the high state data set only.

Note that SSC models are in general not suited to describe the low frequency (i.e.\ radio) emission. Photons of these energies are self-absorbed in the radiation field of the SSC emission region. The observed flux in the radio regime is probably produced by cooled electrons from an outer region of the jet which are unimportant for the modelling of the higher frequencies \citep{MaraschiTavecchio03}.

Taking this into account, both models are in reasonable agreement with the simultaneous data of 1ES\,2344$+$514. The quasi-simultaneous 1FGL points disfavour the one-zone model fits having a $\gamma_\mathrm{min} = 1$. Clearly the fits described by $\gamma_\mathrm{min} = 4000$ are preferred, or in general values being closer to 4000. However, also for the one-zone fits with high $\gamma_\mathrm{min}$, a softer spectral index in the HE regime would improve the compatibility with the 1FGL data. 

The derived model parameters (see Table~\ref{tab:modelling_results}) are in good agreement with typical values for HBLs. Only $\gamma_\mathrm{min}$ seems rather high for the one-zone model, as does $\gamma_\mathrm{max}$ for the low state two-zone model. Considering the exceptional faintness of 1ES\,2344$+$514 across several energy bands, this concordance is not necessarily expected. Consequently, either the low flux state detected in this campaign does (still) not represent the ``quiescent'' state of the source, or the quiescent state model parameters do not differ considerably from the already known ones. The ULs on size and magnetic field strength in the dominating radio emitting region derived from VLBA observations \citep{Sokolovsky_2344_VLBA} do not contradict the parameters of the blazar emission zone as given in Table~\ref{tab:modelling_results}.

\begin{table*}
\caption{Model parameters. See text or reference for an explanation of the models. The used data sets and their simultaneity are discussed in the text.}\label{tab:modelling_results}
\centering
\begin{tabular}{l|cc|cc||cc|c|cc}\hline\hline
Reference & \multicolumn{4}{|c||}{This Campaign} &  \multicolumn{2}{c|}{\citet{MAGIC_2344}} & \citet{Tavecchio10} & \multicolumn{2}{c}{\citet{VERITAS_2344_MWL}}\\
Model Reference & \multicolumn{2}{|c|}{(1)} & \multicolumn{2}{c||}{(2)} & \multicolumn{2}{c|}{(3)} & (1) & \multicolumn{2}{c}{(3)}\\
Flux Level & low & high & low & high & low & high\tablefootmark{b} & \ldots & low & high\\\hline
$B$ [G] & \multicolumn{2}{|c|}{0.07} & 0.05 & 0.09 & 0.10 & 0.08 & 0.1 & 0.09 & 0.03\\
$\delta$ & \multicolumn{2}{|c|}{20} & 26 & 29 & 8 & 15 & 25 & 13 & 20\\
$R_\mathrm{em}$ [$10^{15}$\,cm] & 3 & 4 & 9 & 5 & \multicolumn{2}{c|}{10} & 4 & \multicolumn{2}{c}{10}\\
$R_\mathrm{acc}$ [$10^{13}$\,cm] & \ldots & \ldots & 8 & 13 & \ldots & \ldots & \ldots & \ldots & \ldots\\\hline
$K$\tablefootmark{a} [$10^5\,\mathrm{cm}^{-3}$] & 4.5 & 1.9 & 0.2 & 0.1 & $\sim$\,0.5 & $\sim$\,0.4 & 0.3 & \multicolumn{2}{c}{$\sim$\,0.4}\\
$e_1$ & \multicolumn{2}{|c|}{2.3} & 2.5 & 2.3 & \multicolumn{2}{c|}{2.2} & 2 & 2.5 & 2.3\\
$e_2$ & 3.4 & 3.2 & 3.5 & 3.3 & \multicolumn{2}{c|}{3.2} & 3.2 & \multicolumn{2}{c}{3.2}\\
$\gamma_\mathrm{min}$ & \multicolumn{2}{|c|}{1 or 4000} & 1800 & 550 & $\sim$\,2500 & $\sim$\,1500 & 1 (8000)\tablefootmark{c} & \multicolumn{2}{c}{$\sim$\,200}\\
$\gamma_\mathrm{break}$ [$10^4$] & 5 & 8 & 10 & 3 & \multicolumn{2}{c|}{$\sim$\,15} & 1 & $\sim$\,40 & $\sim$\,50\\
$\gamma_\mathrm{max}$ [$10^6$] & 0.7 & 1.5 & 6.3 & 1.5 & $\sim$\,0.8 & $\sim$\,1.6 & 0.7 & \multicolumn{2}{c}{$\sim$\,2.0}\\\hline
\end{tabular}
\tablefoot{
\tablefoottext{a}{Note that $K$ is given for the acceleration region at $\gamma_\mathrm{min}$ in \citet{Weidinger10}, whereas the value is defined for the emission region at $\gamma_\mathrm{min} = 1$ for the other two models.}
\tablefoottext{b}{See footnote \ref{ftn:MAGIC}.}
\tablefoottext{c}{Fit has been performed on optical data not corrected for the host galaxy. See also footnote \ref{ftn:Tavecchio}.}
\tablebib{
(1)~\citet{MaraschiTavecchio03}; (2) \citet{Weidinger10}; (3) \citet{Krawczynski04}.}
}
\end{table*}

Comparing the high and low state in terms of the one-zone model, the latter is explained by a softer electron spectral index as well as a lower $\gamma_\mathrm{break}$ and $\gamma_\mathrm{max}$. In the case of the two-zone model, the magnetic field drops consistently accompanied by a higher $\gamma_\mathrm{min}$, $\gamma_\mathrm{break}$ and $\gamma_\mathrm{max}$. The parameter changes of each model are best explained by a change in the acceleration properties of the non-thermal electrons, i.e.\ the efficiency of the underlying Fermi processes drops or rises respectively (we recall that $\gamma_\mathrm{break}$ is not computed self-consistently in case of the one-zone model). This behaviour may be caused by the emitting volume leaving a standing feature along the jet \citep[see e.g.][]{MarscherNatureBLLac} or, more likely, due to the observation of two independent blobs.

It is interesting to note that the luminosity of the inverse Compton component of the low (i.e.\ low X-ray flux) state SED exceeds the one of the high state SED for all applied models. Specifically in the case of the one-zone model, this makes the bolometric luminosities $L_{\mathrm{bol}}$ of the two flux states comparable. For $\gamma_{\mathrm{min}} = 1$ (4000), $L_{\mathrm{bol, low}} = 10^{44.7}$ ($10^{44.6}$) and $ L_{\mathrm{bol, high}} = 10^{44.8}$  ($10^{44.7}$) erg\,s$^{-1}$, which is basically identical considering the uncertainties involved.

Due to the differences between the two model approaches at sub-optical frequencies and in the hard X-ray to soft gamma-ray band, it is in principle possible to distinguish between the validity of the models. The first frequency band is covered by \emph{Planck}, though no detection of the source has been reported in The Early Release
Compact Source Catalogue \citep{PlanckEarlyCatalog}, containing the results of the first ten months of operation. \emph{Swift} BAT, \emph{INTEGRAL} IBIS as well as AGILE-GRID and \emph{Fermi}-LAT cover the second window, but are not sensitive enough to detect the source during low flux states on short time scales. To exclude one of these models, a more sensitive instrumentation than currently available is needed.

\begin{figure}
\resizebox{\hsize}{!}{\includegraphics{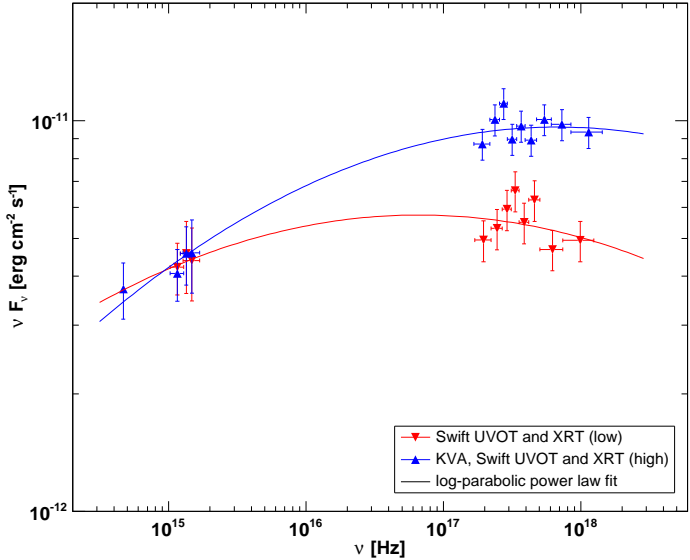}}
\caption{Simultaneous KVA, \emph{Swift} UVOT and \emph{Swift} XRT data from the low (red downward triangles) and high (blue upward triangles) state. The solid lines describe a fit with a log-parabolic power law as stated in Eq.~\ref{eq:logPL}. R-band and UV data have been de-reddened, the former additionally host galaxy corrected. X-ray data have been corrected for Galactic absorption.}\label{fig:2344_synchpeak}
\end{figure}

Within this campaign, a small shift between the synchrotron peak in the high and low state may be present (see Fig.~\ref{fig:2344_SED}). A peak estimate from the data has been obtained by fitting the optical and X-ray SED points by a parabolic power-law in apex form \citep[see e.g.][]{Tramacere07}:
\begin{equation}
\nu F_{\nu} = f_0 \cdot 10^{-b \cdot \left(\log_{10}\left(\nu / \nu_\mathrm{peak}\right)\right)^2}\,\mathrm{erg}\,\mathrm{cm}^{-2}\,\mathrm{s}^{-1}\label{eq:logPL}
\end{equation}
where $\nu_\mathrm{peak}$ is the frequency of the synchrotron peak. The result is shown in Fig.~\ref{fig:2344_synchpeak}. From the fit to the low state data ($\chi^2\mathrm{/d.o.f.} = 6.5/8$), the peak energy is determined marginally significant as $\left(0.27 \pm 0.24\right)$\,keV. In the high state case, the goodness of fit is higher ($\chi^2\mathrm{/d.o.f.} = 5.6/10$), but the fit fails to determine the peak energy ($\left(2.8 \pm 4.5\right)$\,keV). Additionally, a parabolic power law is not clearly preferred over a simple power law in both cases (logarithmic likelihood ratio of 79.5\,\% and 95.7\,\% for the low and high state, respectively). Consequently the data are insufficient to directly constrain changes of the peak energy. As determined from the one-zone $\gamma_\mathrm{min} = 4000$ (two-zone) modelling, the peak shifted from $\sim$\,0.15\,keV to $\sim$\,1.7\,keV ($\sim$\,0.13\,keV to $\sim$\,0.46\,keV) between the low and high flux state. All these values are far from the extreme blazar characteristics 1ES\,2344$+$514 has shown during high X-ray flux states, which is expected due to the rather small flux difference between the two states observed here.

\subsection{Comparison with Archival Campaigns}
We compare the model parameters obtained from this campaign with three other previous MW data sets reported for 1ES\,2344$+$514. The parameters are listed in Table~\ref{tab:modelling_results}.

\subsubsection{Archival Campaign Description}
In \citet{MAGIC_2344}, a homogeneous one-zone SSC model \citep{Krawczynski04} was applied to both a low and high state of the source. The low state data consisted of simultaneous measurements of MAGIC, KVA and an ASM UL taken between 08/2005 and 01/2006, where \emph{BeppoSAX} data taken 06/1998 were added as a low state X-ray spectral template. Data by \emph{BeppoSAX} and a simultaneous UL by Whipple were combined with an archival Whipple high state spectrum to describe the high state\footnote{We note that the Whipple high state fluxes had been adopted incorrectly in \citet{MAGIC_2344}, leading to a softer spectrum with overall lower flux.\label{ftn:MAGIC}}. Consequently the simultaneous data set was not very constraining and the SED models rather speculative, as noted by the authors.

\citet{Tavecchio10} were using the results from the first three months of \emph{Fermi}-LAT observations and combined them with archival, non-contemporaneous data of 1ES\,2344$+$514 from the radio up to the VHE band\footnote{We note that the optical data used for the SED modelling had not been host-galaxy corrected. Increasing $\gamma_\mathrm{min}$ from 1 to $\sim$\,8000 can compensate for the missing correction.\label{ftn:Tavecchio}}. They used the one-zone SSC model also applied in this work.

A more recent campaign, also using the model of \citet{Krawczynski04}, combined measurements by VERITAS, \emph{RXTE} PCA, \emph{Swift} XRT and \emph{Swift} UVOT from 10/2007 until 01/2008 \citep{VERITAS_2344_MWL}. The low state SED was modelled using the time-averaged VERITAS measurements (excluding a large flare) and a representative moderate X-ray flux 1-day spectrum by XRT and PCA, similar to the procedure used in this campaign for the low state SED. Note however that the MAGIC light curve from this campaign did not show significant variability, opposite to the VERITAS measurements even after the exclusion of the flare ($F_\mathrm{var} = \left(34 \pm 16\right)$\,\%). The high state SED data set was built from the highest fluxes measured by VERITAS and \emph{Swift} XRT together with the corresponding UVOT data. Though the gamma-ray and X-ray flares seem to have been correlated, they were separated by $> 24$\,hours. Note that the flux doubling time scale of the VHE and X-ray flare was also $\sim$\,24\,hours, thus the true simultaneous fluxes could have been different by a factor of $\sim$\,2. The model predicted significantly higher, in the high state by $\sim$\,one order of magnitude, fluxes than the 1FGL points. Since these were taken after the MW campaign, higher fluxes in the LAT range during the observations indeed cannot be excluded, despite the rather constant emission of the source for two years since the launch of \emph{Fermi}.

\subsubsection{Model Parameter Comparison}
While the model parameters are in general interdependent and hence difficult to compare, we can more easily investigate general trends. Note that the models shown in \citet{MAGIC_2344} and \citet{Tavecchio10} were not constrained by simultaneous measurements as well as partly affected by incorrectly adopted data (see footnotes \ref{ftn:MAGIC} and \ref{ftn:Tavecchio}), thus the results have to be taken with care. All models indicate a rather weak magnetic field and high Doppler factor, similar to values typically found for HBLs \citep[see e.g.][]{Tavecchio10}. Variability is not explained in a unique way. For instance, changes in $B$ and $\delta$ can occur, but do not have to. Variability mostly arises by changes in the electron distributions, i.e.\ different spectral indices or energy distributions. Particularly in the case of this campaign, the spectral indices are harder in the high flux state. It is interesting to note that the size of the emission region is often not changing between the different flux states presented in Table \ref{tab:modelling_results}. That can be explained within the standing shock scenario, where variability would be caused by structural changes due to variations of the flow \citep{Tagliaferri08}. However, this behaviour is not expected if the two states correspond to the emission of one moving and expanding blob of electrons within the jet \citep[e.g.][]{Atoyan99,Sikora01} at two different times. This argument does not apply in the case of  \citet{MAGIC_2344} though since the presented low and high state are not causally connected but artificially constructed. For \citet{VERITAS_2344_MWL}, another flare occurred between the high and low state SEDs, giving evidence that there is no causal connection. Moreover, the time differences between the two flux states are, also in the case of this campaign, too long to allow both to be caused by the same emitting region, at least if the emission region is visualised as a travelling and expanding (with $v_{\mathrm{exp}}$\,$\sim$\,$c$) blob (which would expand by several orders of magnitudes within one week). To allow for a causal connection between the high and low state, the emission region has to be externally confined in some way (e.g.\ by magnetic fields).

In general, the parameters found in this campaign are in good agreement with archival values. It should however be noted that the range for some parameters is rather broad, due to the inter-dependencies of the parameters and non-unique solutions of the model fits. More data and more MW campaigns in different flux states are needed to fully constrain the models and reduce the allowed parameter ranges.

\section{Summary and Conclusions}\label{conclusion}
In this paper, the results from the first MW campaign on 1ES\,2344$+$514 from the radio to the TeV band have been presented and discussed, also taking into account multi-band long-term data. The MW observations took place from 10/2008 until 01/2009, where the $\sim$\,40\,day long core campaign was conducted in October and November. The source was found at low to modest flux states at radio and optical frequencies, whereas at X-rays and gamma rays, the flux level was amongst the lowest ever reported for 1ES\,2344$+$514. Due to this faintness, HE observations did not result in a detection during the campaign, and the time-averaged VHE detection was only marginally significant. Nevertheless we were able to obtain a reliable VHE spectrum due to the long observation time of $\sim$\,20\,hours, the good event statistics and the source being a well-known VHE emitter.

The VHE analysis suggested a rather hard spectral index, which, if real, would be opposite to the ``harder when brighter'' trend found in general for HBLs. The flux was consistent with constant during the campaign. At X-rays, a moderate flare was detected with $\sim$\,halving of the flux within several days. During the flare, hints for a counter-clockwise behaviour in the hardness ratio\,--\,integral flux plane were found, indicating that the flare was caused by a shock front characterised by comparable cooling and acceleration time scales. This finding was strengthened by the constant spectral index during the flare. Taking into account all X-ray observations during the campaign, the spectral index still did not show significant variability, though a potential correlation between the X-ray index and the X-ray integral flux was visible. The evolution of the hardness ratio with the integral flux corroborated that ``harder when brighter'' trend, confirming findings reported in the literature for 1ES\,2344$+$514. No significant variability could be found at optical and UV frequencies. From VLBA observations, the size of the radio core could be determined or constrained at several frequencies, yielding values of the order of 10$^{17}$\,cm. This is more than one order of magnitude above the size determined from SED modelling, indicating a different origin of the radio and SSC emission.

1ES\,2344$+$514 exhibited significant variability only at low frequency radio and X-ray bands during the campaign. Due to that as well as unfortunate sampling and technical problems, the basis for cross-band correlations for the time of the MW campaign is too short for a meaningful investigation. For a flare observed at the end of the core campaign, indications were found suggesting it may have been caused by injection of fresh electrons into the jet.

On time scales longer than this campaign, significant variability was evident for the radio, optical and X-ray regimes whereas the high energy gamma-ray light curve from 2FGL was consistent with being constant. In contrast to the low and constant emission found by \emph{Fermi}-LAT, the events with energies above 100\,GeV detected from 1ES\,2344$+$514 indicate that the source may have a comparably high flaring duty cycle.

Different feature characteristics were found in the Effelsberg light curve at low radio frequencies, indicating a possible re-acceleration of particles within the jet. The difference between these features may be explained by changes in the environment of the particles. The behaviour of the combined long-term radio spectra of the source gave rise to interpreting the emission as a two-component system composed of quiescent diffuse emission overlaid by frequent outbursts. The signature of such shocks should be traceable from higher to lower radio frequencies. One flaring event at 37\,GHz was visible on the investigated time scales, but the expected signatures could not be found in the other radio bands. The OVRO measurements had some gaps during these days but did not show evidence of a significant flux increase. Two of the nine photons with an energy $> 100$\,GeV were detected by \emph{Fermi}-LAT around the time of that flare. A counterpart of this event was not found in the optical R-band despite rather good optical coverage, which would hint at a different emission region of the 37\,GHz and R-band emission if not due to sampling effects. The flare on its own represents a rare event for HBLs concerning its amplitude and time scale. A long-term correlation analysis between the 15.0\,GHz and R-band was conducted, yielding no significant correlation between the two bands.

The observed flat parsec-scale radio spectrum together with the frequency-dependent core size (Table~\ref{tab:coresize}) and position shift, indicated by a comparison of core--jet component distances measured at 15.4 and 43.2\,GHz by \citet{Piner04} and \citet{Piner10}, can be interpreted as a signature of a \citet{BlandfordKoenigl79} type jet. Analysing all MOJAVE observations conducted until today, no significant motion of the three identified jet components on time scales of eleven years could be found, opposite to claims in previous publications. The apparent jet speeds of the components were $\beta_\mathrm{app} < 0.13$, with the most constraining value having been $-0.01 \pm 0.02$.

Monitoring at soft and hard X-rays revealed only one significant individual detection, though a general trend of positive flux for BAT and ISGRI was apparent, leading to the detection of the source by BAT from 58\,months of data. The individual detection, found by ASM, was coincident within a few days with a higher state seen by XRT and a hint for an R-band flare seen by KVA. Also the Effelsberg measurements showed increased activity around this time period. However the sampling was insufficient for a meaningful investigation of the origin of the flare. The BAT light curve was significantly variable. The long-term trend measured by ASM did not show a hint of a positive signal.

From the observations, (quasi-)simultaneous SEDs for a low and high X-ray flux state were constructed and modelled using a one-zone SSC as well as a self-consistent two-zone SSC model. Both could describe the data well, however quasi-simultaneous HE data posed some challenges for the modelling. In particular, these disfavoured the one-zone models having a $\gamma_{\mathrm{min}}$ of unity, being in general better described by the upper part of the tested parameter range. The one- and two-zone models suggested a shift of the first SED peak by $\sim$\,1.1 and $\sim$\,0.4 orders of magnitude, respectively. Direct fitting of the combined optical and X-ray data did not result in a firm determination of the peak energies. The individual parameters retrieved from the one- and two-zone modelling were mostly in agreement between these two different model approaches for each of the two flux states. They were consistent with values found in archival campaigns as well as standard parameter ranges for HBLs. This concordance is not self-evident in the context of ``quiescent'' state emission, where the quiescent spectrum should be dominated by a low and constant flux component which possibly has different spectral characteristics. Either the ``quiescent'' state was not detected within this MW campaign, or the corresponding model parameters do not differ significantly from the typical values. The two applied models showed significant differences at high radio frequencies and in the hard X-ray to HE bands. In the future, instruments more sensitive in these regimes could probe the validity of the models.

\begin{acknowledgements}
We would like to thank the anonymous referee for constructive comments.
The MAGIC Collaboration would like to thank the Instituto de Astrof\'{\i}sica de
Canarias for the excellent working conditions at the
Observatorio del Roque de los Muchachos in La Palma.
The support of the German BMBF and MPG, the Italian INFN, 
the Swiss National Fund SNF, and the Spanish MICINN is 
gratefully acknowledged. This work was also supported by the CPAN CSD2007-00042 and MultiDark
CSD2009-00064 projects of the Spanish Consolider-Ingenio 2010
programme, by grant 127740 of 
the Academy of Finland,
by the DFG Cluster of Excellence ``Origin and Structure of the 
Universe'', by the DFG Collaborative Research Centers SFB823/C4 and SFB876/C3,
and by the Polish MNiSzW grant 745/N-HESS-MAGIC/2010/0.

The AGILE Mission is funded by the Italian Space Agency (ASI), with scientific and programmatic participation by the Italian Institute of Astrophysics (INAF) and the Italian Institute of Nuclear Physics (INFN). Research partially supported through the ASI grants no.\ I/089/06/2 and I/042/10/0.

The \emph{Fermi}-LAT Collaboration acknowledges support from a number of agencies and institutes for both development and the operation of the LAT as well as scientific data analysis. These include NASA and DOE in the United States, CEA/Irfu and IN2P3/CNRS in France, ASI and INFN in Italy, MEXT, KEK, and JAXA in Japan, and the K.\,A.\,Wallenberg Foundation, the Swedish Research Council and the National Space Board in Sweden. Additional support from INAF in Italy and CNES in France for science analysis during the operations phase is also gratefully acknowledged.

We gratefully acknowledge the entire \emph{Swift} team, the duty scientists and science planners for the dedicated support, making these observations possible.

This research has made use of the XRT Data Analysis Software (XRTDAS) developed under the responsibility of the ASI Science Data Center (ASDC), Italy.

This research is partly based on observations with the 100-m telescope of the MPIfR (Max-Planck-Institut f\"ur Radioastronomie) at Effelsberg and has made use of observations with the IRAM 30-m telescope. IRAM is supported by INSU/CNRS (France), MPG (Germany) and IGN (Spain).

The Mets\"ahovi team acknowledges the support from the Academy of Finland to our observing projects (numbers 212656, 210338, 121148, and others).

This research has made use of data from the MOJAVE database that is maintained by the MOJAVE team \citep{Lister09_V}. The MOJAVE project is supported under NASA-\emph{Fermi} grant NNX08AV67G.

The VLBA is a facility of the National Science Foundation operated by the National Radio Astronomy Observatory under cooperative agreement with Associated Universities, Inc. This work made use of the Swinburne University of Technology software correlator, developed as part of the Australian Major National Research Facilities Programme and operated under licence.

RATAN-600 operations were carried out with the financial support of the Ministry of Education and Science of the Russian Federation (contracts 16.518.11.7062 and 16.552.11.7028).

This publication makes use of data products from the Two Micron All Sky Survey, which is a joint project of the University of Massachusetts and the Infrared Processing and Analysis Center/California Institute of Technology, funded by the National Aeronautics and Space Administration and the National Science Foundation.

This research has made use of NASA's Astrophysics Data System.

Part of this work is based on archival data, software or on-line services provided by the ASI Science Data Center (ASDC).

This research has made use of the NASA/IPAC Extragalactic Database (NED) which is operated by the Jet Propulsion Laboratory, California Institute of Technology, under contract with the National Aeronautics and Space Administration.

This research has made use of SAOImage DS9, developed by Smithsonian Astrophysical Observatory.

This research has made use of the SIMBAD database, operated at CDS, Strasbourg, France.

Y.\ A.\ K., Y.\ Y.\ K., and K.\ V.\ S.\ were supported in part by the Russian Foundation for Basic Research (projects 11-02-00368, 12-02-33101), the basic research program ``Active processes in galactic and extragalactic objects'' of the Physical Sciences Division of the Russian Academy of Sciences, and the Ministry of Education and Science of the Russian Federation (agreement No.~8405). Y.\ Y.\ K.\ was also supported by the Dynasty Foundation.

\end{acknowledgements}

\bibliographystyle{aa}
\bibliography{bibfile}

\appendix
\section{Detailed Results}
\begin{sidewaystable*}
\centering
\rule{0.\textheight}{0.745\textheight}
\caption{\emph{Swift} XRT results.}\label{tab:Swift_XRT_results}
\begin{tabular}{lll|lll|llllll|l}
\hline\hline
& & & \multicolumn{3}{c|}{Simple Power Law Fit} & \multicolumn{6}{c|}{Log-Parabolic Power Law Fit} &\\
Obs.\ ID\tablefootmark{a} & MJD$_\mathrm{start}$ & Exp.\tablefootmark{b} & $F \left(2-10\,\mathrm{keV}\right)$\tablefootmark{c} & $a$\tablefootmark{d} & $\chi^2_{\mathrm{red}}\mathrm{/d.o.f.}$ & $F \left(0.2-1\,\mathrm{keV}\right)$\tablefootmark{e} & $F \left(2-10\,\mathrm{keV}\right)$\tablefootmark{e} & $a$\tablefootmark{f} & $b$\tablefootmark{f} & $\chi^2_{\mathrm{red}}\mathrm{/d.o.f.}$ & L\tablefootmark{g} & HR\tablefootmark{h}\\
& & [ks] & [$10^{-12}\,\mathrm{erg}\,\mathrm{cm}^{-2}\,\mathrm{s}^{-1}$] & & &  [$10^{-12}\,\mathrm{erg}\,\mathrm{cm}^{-2}\,\mathrm{s}^{-1}$] & [$10^{-12}\,\mathrm{erg}\,\mathrm{cm}^{-2}\,\mathrm{s}^{-1}$] & & & & [\%] &\\\hline
35031019 & 54730.158 & 0.59 & \ldots\tablefootmark{i} & \ldots\tablefootmark{i} & \ldots\tablefootmark{i} & \ldots\tablefootmark{i} & \ldots\tablefootmark{i} & \ldots\tablefootmark{i} & \ldots\tablefootmark{i} & \ldots\tablefootmark{i} & \ldots\tablefootmark{i} &\ldots\tablefootmark{i}\\
35031021 & 54745.554 & 1.75 & 10.8 $ \pm $ 0.9 & 2.03 $ \pm $ 0.11 & 0.77/17 & 2.50 $ \pm $ 0.30 & \phantom{0}9.2 $ \pm $ 0.8 & 1.79 $ \pm $ 0.25 & 0.51 $ \pm $ 0.30 & 0.61/18 & 85.1 & $1.17 \pm 0.12$\\
35031022 & 54749.513 & 0.94 & \phantom{0}9.8 $ \pm $ 1.4 & 2.12 $ \pm $ 0.23 & 0.79/7 & 3.05 $ \pm $ 0.50 & \phantom{0}9.8 $ \pm $ 1.5 & 2.12 $ \pm $ 0.35 & 0.00 $ \pm $ 0.00 & 0.93/6 & \phantom{0}0.0 & $0.80 \pm 0.12$\\
35031023 & 54757.762 & 1.23 & \phantom{0}9.5 $ \pm $ 1.1 & 2.14 $ \pm $ 0.18 & 1.55/13 & 3.09 $ \pm $ 0.50 & \phantom{0}9.5 $ \pm $ 1.4 & 2.14 $ \pm $ 0.30 & 0.00 $ \pm $ 0.00 & 1.68/12 & \phantom{0}0.0 & $0.95 \pm 0.13$\\
35031024 & 54759.895 & 2.23 & 14.2 $ \pm $ 0.9 & 1.94 $ \pm $ 0.09 & 1.10/30 & 2.90 $ \pm $ 0.25 & 13.0 $ \pm $ 0.8 & 1.80 $ \pm $ 0.18 & 0.27 $ \pm $ 0.25 & 1.07/29 & 82.9 & $1.21 \pm 0.10$\\
35031025 & 54760.899 & 2.31 & 14.4 $ \pm $ 0.8 & 1.98 $ \pm $ 0.08 & 1.04/42 & 3.16 $ \pm $ 0.22 & 13.0 $ \pm $ 0.9 & 1.80 $ \pm $ 0.16 & 0.36 $ \pm $ 0.26 & 0.95/41 & 96.0 & $1.17 \pm 0.09$\\
35031026 & 54761.904 & 2.27 & 13.4 $ \pm $ 0.7 & 1.97 $ \pm $ 0.08 & 1.32/32 & 2.91 $ \pm $ 0.23 & 12.3 $ \pm $ 0.9 & 1.84 $ \pm $ 0.17 & 0.26 $ \pm $ 0.25 & 1.29/31 &85.6 & $1.22 \pm 0.10$\\
35031027 & 54762.908 & 2.42 & 10.9 $ \pm $ 0.6 & 2.02 $ \pm $ 0.09 & 1.12/28 & 1.67 $ \pm $ 0.22 & 10.3 $ \pm $ 0.8 & 1.94 $ \pm $ 0.16 & 0.17 $ \pm $ 0.17 & 1.12/27 & 69.9 & $1.06 \pm 0.09$\\
35031028 & 54763.167 & 4.91 & 10.5 $ \pm $ 0.6 & 2.03 $ \pm $ 0.06 & 1.20/48 & 2.61 $ \pm $ 0.17 & \phantom{0}9.7 $ \pm $ 0.6 & 1.92 $ \pm $ 0.12 & 0.25 $ \pm $ 0.21 & 1.15/47 & 93.0 & $1.07 \pm 0.07$\\
35031029 & 54764.857 & 1.58 & 10.0 $ \pm $ 0.8 & 2.09 $ \pm $ 0.12 & 0.97/17 & 2.64 $ \pm $ 0.30 & \phantom{0}9.0 $ \pm $ 1.0 & 1.92 $ \pm $ 0.27 & 0.35 $ \pm $ 0.34 & 0.93/16 & 78.5 & $1.09 \pm 0.12$\\
35031030 & 54765.917 & 2.53 & \phantom{0}9.6 $ \pm $ 0.7 & 2.01 $ \pm $ 0.08 & 1.36/26 & 2.11 $ \pm $ 0.23 & \phantom{0}8.4 $ \pm $ 0.8 & 1.75 $ \pm $ 0.19 & 0.51 $ \pm $ 0.30 & 1.07/25 & 98.6 & $1.13 \pm 0.10$\\
35031031 & 54766.865 & 1.19 & \phantom{0}8.1 $ \pm $ 1.1& 2.10 $ \pm $ 0.20 & 0.88/8 & 2.25 $ \pm $ 0.50 & \phantom{0}7.2 $ \pm $ 1.2 & 1.99 $ \pm $ 0.40 & 0.26 $ \pm $ 0.26 & 0.99/7 & 24.9 & $0.86 \pm 0.13$\\
35031032 & 54767.869 & 2.75 & \phantom{0}8.0 $ \pm $ 0.6 & 2.15 $ \pm $ 0.10 & 0.76/26 & 2.58 $ \pm $ 0.25 & \phantom{0}8.0 $ \pm $ 0.9 & 2.14 $ \pm $ 0.13 & 0.00 $ \pm $ 0.00 & 0.79/25 & \phantom{0}6.9 & $0.87 \pm 0.08$\\
35031033 & 54768.806 & 2.29 & \phantom{0}7.9 $ \pm $ 0.5 & 2.04 $ \pm $ 0.11 & 0.99/18 & 1.82 $ \pm $ 0.20 & \phantom{0}6.9 $ \pm $ 0.8 & 1.78 $ \pm $ 0.26 & 0.51 $ \pm $ 0.42 & 0.84/17 & 92.9 & $1.13 \pm 0.12$\\
35031034 & 54769.932 & 1.87 & 11.5 $ \pm $ 0.7 & 1.96 $ \pm $ 0.10 & 1.09/20 & 2.44 $ \pm $ 0.25 & 10.7 $ \pm $ 0.9 & 1.83 $ \pm $ 0.22 & 0.25 $ \pm $ 0.25 & 1.07/19 & 76.4 & $1.21 \pm 0.12$\\
35031035 & 54770.881 & 2.10 & \phantom{0}9.6 $ \pm $ 0.7 & 2.06 $ \pm $ 0.11 & 0.58/21 & 2.60 $ \pm $ 0.25 & \phantom{0}9.3 $ \pm $ 1.0 & 2.02 $ \pm $ 0.20 & 0.00 $ \pm $ 0.00 & 0.60/20 & 31.8 & $1.01 \pm 0.10$\\
35031036 & 54771.933 & 1.65 & \phantom{0}9.9 $ \pm $ 0.8 & 1.99 $ \pm $ 0.11 & 1.36/15 & 1.93 $ \pm $ 0.22 & \phantom{0}8.6 $ \pm $ 0.9 & 1.58 $ \pm $ 0.32 & 0.73 $ \pm $ 0.50 & 1.01/14 & 97.7 & $1.23 \pm 0.14$\\
35031037 & 54772.892 & 0.19 & \ldots\tablefootmark{i} & \ldots\tablefootmark{i} & \ldots\tablefootmark{i} & \ldots\tablefootmark{i} & \ldots\tablefootmark{i} & \ldots\tablefootmark{i} & \ldots\tablefootmark{i} & \ldots\tablefootmark{i} & \ldots\tablefootmark{i} &\ldots\tablefootmark{i}\\
35031038 & 54773.892 & 0.94 & 13.2 $ \pm $ 1.7 & 1.86 $ \pm $ 0.22 & 0.30/7 & 2.43 $ \pm $ 0.50 & 12.4 $ \pm $ 2.0 & 1.80 $ \pm $ 0.50 & 0.14 $ \pm $ 0.14 & 0.35/6 & \phantom{0}0.0 & $1.17 \pm 0.18$\\
35031039 & 54777.483 & 1.59 & 10.7 $ \pm $ 0.7 & 2.16 $ \pm $ 0.12 & 1.36/22 & 3.55 $ \pm $ 0.30 & 10.6 $ \pm $ 1.2 & 2.15 $ \pm $ 0.25 & 0.00 $ \pm $ 0.00 & 1.43/19 & 89.2 & $0.99 \pm 0.10$\\
35031040 & 54784.592 & 1.06 & 16.6 $ \pm $ 1.7 & 1.76 $ \pm $ 0.13 & 0.32/13 & 2.48 $ \pm $ 0.30 & 15.9 $ \pm $ 1.9 & 1.69 $ \pm $ 0.30 & 0.13 $ \pm $ 0.13 & 0.33/12 & 33.5 & $1.47 \pm 0.18$\\\hline
\end{tabular}
\tablefoot{
\tablefoottext{a}{\emph{Swift} observation ID.}
\tablefoottext{b}{\emph{Swift} XRT exposure.}
\tablefoottext{c}{Integral flux between 2 and 10\,keV determined by a simple power law fit from 0.3\,--\,10\,keV.}
\tablefoottext{d}{Spectral index determined by a simple power law fit from 0.3\,--\,10\,keV.}
\tablefoottext{e}{Integral flux determined by a log parabolic power law fit from 0.3\,--\,10\,keV.}
\tablefoottext{f}{Spectral indices determined from a log-parabola fit from 0.3\,--\,10\,keV.}
\tablefoottext{g}{Probability that the log-parabolic power law fit is preferred over the simple power law fit by means of a logarithmic likelihood ratio test.}
\tablefoottext{h}{Hardness ratio, defined here as $\mathrm{counts}_{\left(2-10\,\mathrm{keV}\right)}/\mathrm{counts}_{\left(0.2-1\,\mathrm{keV}\right)}$.}
\tablefoottext{i}{Observation time too short for extracting results.}
}
\end{sidewaystable*}

\clearpage

\begin{table}[h]
\caption{Analysed MAGIC data sets and results using detection cuts to determine the significance and open cuts for the integral fluxes.}\label{tab:MAGIC_results}
\centering
\begin{tabular}{lllll}
\hline\hline
Data Set\tablefootmark{a} & Obs.\ Time\tablefootmark{b} & $t_\mathrm{eff}$\tablefootmark{c} & $S$\tablefootmark{d} & $F \left(> 170\,\mathrm{GeV}\right)$\tablefootmark{e}\\
& [MJD] & [h] & & [$10^{-12}\,\mathrm{ph}\,\mathrm{cm}^{-2}\,\mathrm{s}^{-1}$]\\\hline
all data & 54780.419 & 20.75 & \phantom{$-$}3.5 & \phantom{0}$7.4 \pm \phantom{0}2.1$\\
period 1 & 54763.433 & 10.26 & \phantom{$-$}1.9 & \phantom{0}$3.2 \pm \phantom{0}2.9$\\
period 2 & 54793.858 & 10.49 & \phantom{$-$}3.1 & $11.5 \pm \phantom{0}2.9$\\\hline
21 Oct.\ & 54759.973 & \phantom{0}1.42 & \phantom{$-$}1.3 & $10.6 \pm \phantom{0}8.0$\\
22 Oct.\ & 54760.960 & \phantom{0}1.41 &\phantom{$-$}0.6 & \phantom{0}$9.3 \pm \phantom{0}7.8$\\
23 Oct.\ & 54761.954 & \phantom{0}1.35 & \phantom{$-$}1.1 & $< 17.4$\\
24 Oct.\ & 54762.931 & \phantom{0}1.42 & \phantom{$-$}0.4 & $< 12.5$\\
25 Oct.\ & 54763.949 & \phantom{0}3.22 & \phantom{$-$}1.5 & $< 15.6$\\
26 Oct.\ & 54764.946 & \phantom{0}1.14 & \phantom{$-$}0.0 & $< 19.5$\\
28 Oct.\ & 54766.919 & \phantom{0}0.30 & $-2.0$ & $< \phantom{0}1.7$\\
17 Nov.\ & 54786.862 & \phantom{0}1.03 & \phantom{$-$}2.5 & $27.0 \pm \phantom{0}9.9$\\
18 Nov.\ & 54787.873 & \phantom{0}2.14 & $-0.1$ & $< 19.7$\\
19 Nov.\ & 54788.888 & \phantom{0}1.77 & \phantom{$-$}2.0 & $16.0 \pm \phantom{0}7.2$\\
24 Nov.\ & 54793.866 & \phantom{0}2.12 & \phantom{$-$}1.2 & \phantom{0}$9.5 \pm \phantom{0}6.6$\\
25 Nov.\ & 54794.866 & \phantom{0}2.01 & \phantom{$-$}1.4 & \phantom{0}$6.5 \pm \phantom{0}6.4$\\
28 Nov.\ & 54797.841 & \phantom{0}0.62 & $-1.1$ & $< 16.5$\\
01 Dec.\ & 54800.879 & \phantom{0}0.81 & \phantom{$-$}2.0 & $22.3 \pm 10.0$\\\hline
\end{tabular}
\tablefoot{
\tablefoottext{a}{If dates are given, they correspond to the day following the observation night.}
\tablefoottext{b}{Arithmetic average of observation duration.}
\tablefoottext{c}{Effective observation time.}
\tablefoottext{d}{Significance of the signal calculated according to \citet{LiMa83} Eq.~17.}
\tablefoottext{e}{Measured integral flux. ULs are given with 95\,\% c.l.. We recall that the fluxes and signal significances were determined using different cuts.}
}
\end{table}

\begin{table}
\caption{Calculated HE luminosities and number of events above 100\,GeV detected by \emph{Fermi}-LAT from five HBLs.}\label{tab:FermiHE_comp}
\centering
\begin{tabular}{llllll}
\hline\hline
Source & z\tablefootmark{a} & Index\tablefootmark{b} & $L_{\mathrm{60\,GeV}}$\tablefootmark{c} & $N$\tablefootmark{d} & $N_{\mathrm{s}}$\tablefootmark{e}\\
& & & [$10^{43}$\,erg\,s$^{-1}$] & &\\\hline
Mrk\,421 & 0.030 & $1.77 \pm 0.01$ & $19.2 \pm 1.1$ & 35 & 18\\
Mrk\,501 & 0.034 & $1.74 \pm 0.03$ & $6.95 \pm 0.75$ & 16 & 10\\
1ES\,2344$+$514 & 0.044 & $1.72 \pm 0.08$ & $2.30 \pm 0.56$ & \phantom{0}9 & \phantom{0}9\\
Mrk\,180 & 0.046 & $1.74 \pm 0.08$ & $2.01 \pm 0.51$ & \phantom{0}1 & \phantom{0}1\\
1ES\,1959$+$650 & 0.048 & $1.94 \pm 0.03$ & $6.00 \pm 0.80$ & \phantom{0}3 & \phantom{0}4\\\hline
\end{tabular}
\tablefoot{
\tablefoottext{a}{Redshift.}
\tablefoottext{b}{Simple power law spectral index measured by \emph{Fermi}-LAT \citep{Fermi_2FGL}. Note that for all sources the simple power law is clearly preferred over a curved description of the spectrum.}
\tablefoottext{c}{Luminosity at 60\,GeV, determined on the basis of the 10\,--\,100\,GeV photon counts reported in \citet{Fermi_2FGL}.}
\tablefoottext{d}{Number of events above 100\,GeV.}
\tablefoottext{e}{Number of events above 100\,GeV scaled to the distance of 1ES\,2344$+$514.}
}
\end{table}

\begin{figure}[!p]
\resizebox{\hsize}{!}{\includegraphics{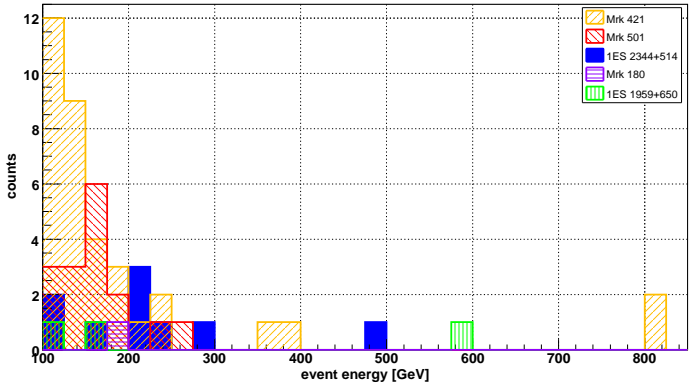}}
\caption{Distribution of events with energies $> 100$\,GeV detected by \emph{Fermi}-LAT from Mrk\,421, Mrk\,501, 1ES\,2344$+$514, Mrk\,180 and 1ES\,1959$+$650.}\label{fig:FermiHE_comp}
\end{figure}

\begin{table*}
\caption{\emph{Swift} UVOT results.}\label{tab:Swift_UVOT_results}
\centering
\begin{tabular}{lllllllll}
\hline\hline
Obs.\ ID\tablefootmark{a} & MJD$_\mathrm{start}$ & Exp.\tablefootmark{b} & $F_\mathrm{V}$ & $F_\mathrm{B}$ & $F_\mathrm{U}$ & $F_\mathrm{UVW1}$ & $F_\mathrm{UVM2}$ & $F_\mathrm{UVW2}$\\
& & [ks] & [mag] & [mag] & [mag] & [mag] & [mag] & [mag]\\\hline
35031019 & 54730.158 & 0.89 & 15.45 $ \pm $ 0.10 & 16.47 $ \pm $ 0.10 & 16.69 $ \pm $ 0.12 & 17.27 $ \pm $ 0.20 & 17.76 $ \pm $ 0.30 & 17.76 $ \pm $ 0.15\\ 
35031021 & 54745.554 & 1.63 & 15.46 $ \pm $ 0.10 & 16.49 $ \pm $ 0.10 & 16.57 $ \pm $ 0.12 & 17.20 $ \pm $ 0.20 & 17.79 $ \pm $ 0.30 & 17.63 $ \pm $ 0.20\\
35031022 & 54749.513 & 0.88 & 15.47 $ \pm $ 0.07 & 16.45 $ \pm $ 0.07 & 16.53 $ \pm $ 0.10 & 17.15 $ \pm $ 0.10 & 17.57 $ \pm $ 0.15 & 17.41 $ \pm $ 0.10\\
35031023 & 54757.769 & 1.14 & 15.46 $ \pm $ 0.10 & 16.52 $ \pm $ 0.10 & 16.71 $ \pm $ 0.12 & 17.30 $ \pm $ 0.20 & 17.74 $ \pm $ 0.30 & 17.49 $ \pm $ 0.15\\
35031024 & 54759.895 & 2.18 & 15.47 $ \pm $ 0.07 & 16.51 $ \pm $ 0.07 & 16.66 $ \pm $ 0.10 & 17.31 $ \pm $ 0.10 & 17.73 $ \pm $ 0.15 & 17.66 $ \pm $ 0.10\\
35031025 & 54760.899 & 2.26 & 15.46 $ \pm $ 0.07 & 16.52 $ \pm $ 0.07 & 16.69 $ \pm $ 0.10 & 17.27 $ \pm $ 0.10 & 17.66 $ \pm $ 0.15 & 17.62 $ \pm $ 0.10\\
35031026 & 54761.904 & 2.23 & 15.46 $ \pm $ 0.07 & 16.47 $ \pm $ 0.07 & 16.56 $ \pm $ 0.10 & 17.16 $ \pm $ 0.10 & 17.68 $ \pm $ 0.15 & 17.66 $ \pm $ 0.10\\
35031027 & 54762.908 & 2.38 & 15.45 $ \pm $ 0.07 & 16.46 $ \pm $ 0.07 & 16.52 $ \pm $ 0.10 & 17.17 $ \pm $ 0.10 & 17.41 $ \pm $ 0.15 & 17.54 $ \pm $ 0.10\\
35031028 & 54763.167 & 4.83 & 15.46 $ \pm $ 0.05 & 16.51 $ \pm $ 0.05 & 16.57 $ \pm $ 0.10 & 17.19 $ \pm $ 0.10 & 17.59 $ \pm $ 0.15 & 17.57 $ \pm $ 0.10\\
35031029 & 54764.857 & 1.52 & 15.50 $ \pm $ 0.07 & 16.46 $ \pm $ 0.10 & 16.50 $ \pm $ 0.10 & 17.10 $ \pm $ 0.15 & 17.33 $ \pm $ 0.20 & 17.50 $ \pm $ 0.10\\
35031030 & 54765.917 & 2.48 & 15.41 $ \pm $ 0.07 & 16.50 $ \pm $ 0.07 & 16.57 $ \pm $ 0.10 & 17.31 $ \pm $ 0.10 & 17.58 $ \pm $ 0.15 & 17.62 $ \pm $ 0.10\\
35031031 & 54766.865 & 1.15 & 15.41 $ \pm $ 0.10 & 16.49 $ \pm $ 0.10 & 16.61 $ \pm $ 0.10 & 17.20 $ \pm $ 0.10 & 17.56 $ \pm $ 0.30 & 17.55 $ \pm $ 0.15\\
35031032 & 54767.869 & 2.68 & 15.48 $ \pm $ 0.07 & 16.51 $ \pm $ 0.10 & 16.65 $ \pm $ 0.10 & 17.37 $ \pm $ 0.10 & 17.80 $ \pm $ 0.20 & 17.78 $ \pm $ 0.10\\
35031033 & 54768.806 & 1.25 & 15.44 $ \pm $ 0.07 & 16.50 $ \pm $ 0.07 & 16.60 $ \pm $ 0.10 & 17.23 $ \pm $ 0.10 & 17.66 $ \pm $ 0.20 & 17.67 $ \pm $ 0.10\\
35031034 & 54769.932 & 1.82 & 15.50 $ \pm $ 0.07 & 16.52 $ \pm $ 0.07 & 16.67 $ \pm $ 0.10 & 17.48 $ \pm $ 0.10 & 17.62 $ \pm $ 0.15 & 17.70 $ \pm $ 0.10\\
35031035 & 54770.881 & 2.05 & \ldots                       & \ldots                      & \ldots                      & \ldots                       & \ldots                       & \ldots\\
35031036 & 54771.933 & 1.63 & 15.49 $ \pm $ 0.07 & 16.51 $ \pm $ 0.07 & 16.73 $ \pm $ 0.10 & 17.38 $ \pm $ 0.10 & 17.66 $ \pm $ 0.15 & 17.58 $ \pm $ 0.10\\
35031037 & 54772.892 & 0.18 & \ldots                       & 16.54 $ \pm $ 0.07 & 16.77 $ \pm $ 0.10 & 17.22 $ \pm $ 0.10 & \ldots                      & \ldots\\
35031038 & 54773.892 & 0.91 & 15.40 $ \pm $ 0.07 & 16.44 $ \pm $ 0.07 & 16.54 $ \pm $ 0.10 & 17.12 $ \pm $ 0.10 & 17.51 $ \pm $ 0.20 & 17.49 $ \pm $ 0.15\\
35031039 & 54777.483 & 1.54 & 15.39 $ \pm $ 0.07 & 16.48 $ \pm $ 0.07 & 16.54 $ \pm $ 0.10 & 17.21 $ \pm $ 0.10 & 17.62 $ \pm $ 0.15 & 17.59 $ \pm $ 0.10\\
35031040 & 54784.592 & 1.03 & 15.50 $ \pm $ 0.07 & 16.52 $ \pm $ 0.07 & 16.69 $ \pm $ 0.10 & 17.36 $ \pm $ 0.10 & 17.52 $ \pm $ 0.15 & 17.68 $ \pm $ 0.10\\\hline
\end{tabular}
\tablefoot{
\tablefoottext{a}{\emph{Swift} observation ID.}
\tablefoottext{b}{\emph{Swift} total exposure of all UVOT filters.}
}
\end{table*}

\begin{figure*}[h]
\resizebox{\hsize}{!}{\includegraphics{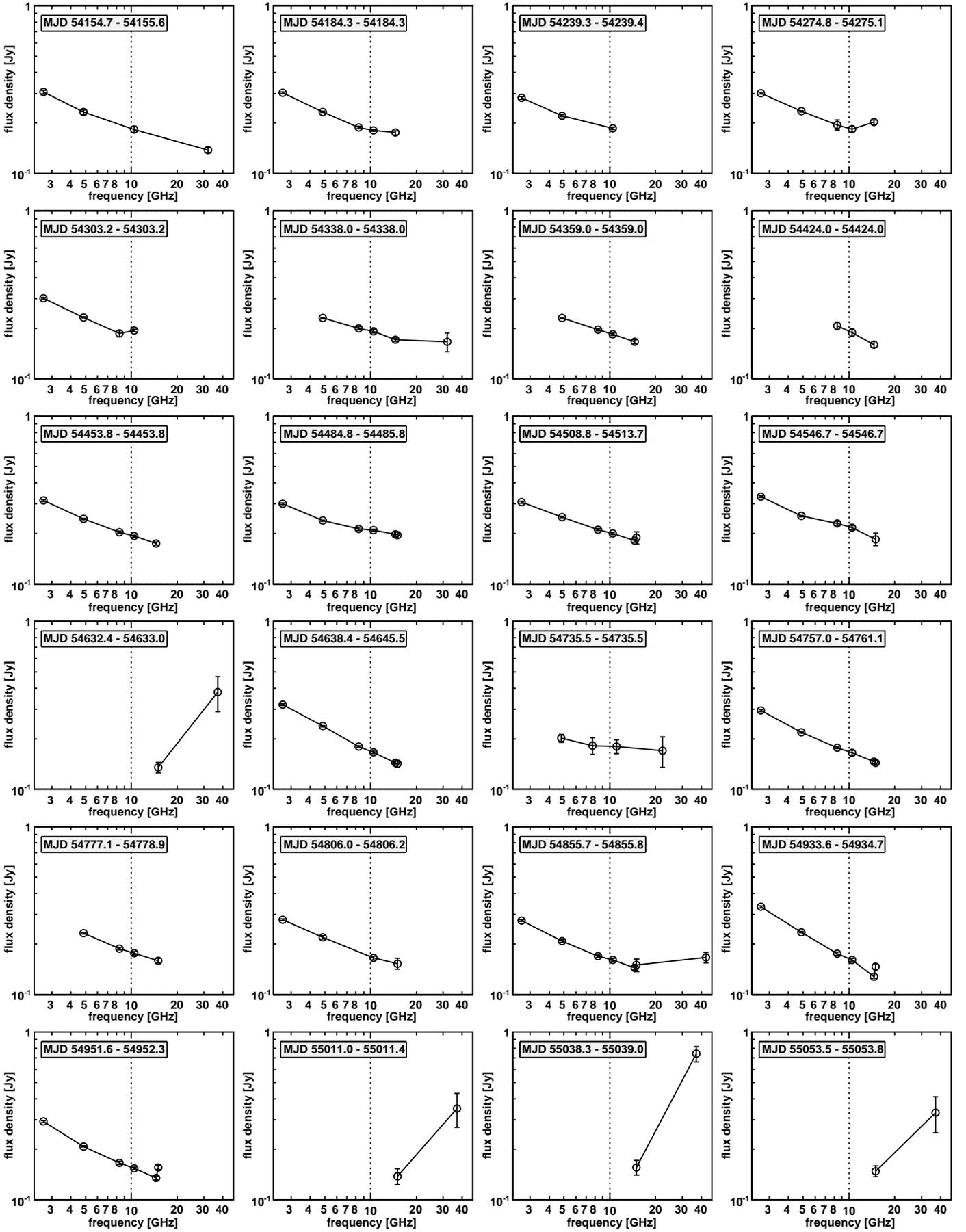}}
\caption{Individual radio spectra of Fig.~\ref{fig:2344_radio_spectra}.}\label{fig:2344_radio_spectra_single}
\end{figure*}

\begin{table*}[h]
\caption{Gaussian Component Properties of 1ES\,2344$+$514 measured with the VLBA at 15.4\,GHz.}\label{tab:jet_components}
\centering
\begin{tabular}{lllllllllll}
\hline\hline
Component & Epoch & $r$\tablefootmark{a} & $\theta$\tablefootmark{b} & $S$\tablefootmark{c} & Maj.\tablefootmark{d} & Axial & PA\tablefootmark{f} & log $T_\mathrm{b}$\tablefootmark{g} & $v_\mathrm{app}$\tablefootmark{h} & $\beta_\mathrm{app}$\tablefootmark{i}\\
& & [mas] & [deg] & [mJy] & [mas] & Ratio\tablefootmark{e} & [deg] & [K] & [$\mu\mathrm{as\,yr^{-1}}$] &\\\hline
\multirow{13}*{Core} & 1999.75 & \ldots & \ldots & 128.9 & 0.05 & 1.00 &  \ldots & 11.37 & \multirow{13}*{\ldots} & \multirow{13}*{\ldots}\\
& 1999.85 & \ldots & \ldots & 136.4 & \ldots & \ldots & \ldots & \ldots & &\\
& 2000.02 & \ldots & \ldots & 126.9 & \ldots & \ldots & \ldots & \ldots & &\\
& 2000.22 & \ldots & \ldots & 134.4 & \ldots & \ldots & \ldots & \ldots & &\\
& 2008.41 & \ldots & \ldots & 107.7 & 0.05 & 1.00 & \ldots & 11.37 & &\\
& 2008.76 & \ldots & \ldots & \phantom{0}97.9 & 0.16 & 0.49 & 322 & 10.59 & &\\
& 2008.81 & \ldots & \ldots & 102.2 & 0.07 & 1.00 & \ldots & 10.98 & &\\
& 2009.15 & \ldots & \ldots & \phantom{0}78.9 & 0.22 & 0.21 & 317 & 10.62 & &\\
& 2009.42 & \ldots & \ldots & \phantom{0}83.6 & 0.05 & 1.00 & \ldots & 11.24 & &\\
& 2009.51 & \ldots & \ldots & \phantom{0}94.3 & 0.17 & 0.27 & 322 & 10.81 & &\\
& 2009.63 & \ldots & \ldots & \phantom{0}81.7 & 0.21 & 0.30 & 316 & 10.50 & &\\
& 2009.94 & \ldots & \ldots & \phantom{0}98.7 & 0.17 & 0.28 & 317 & 10.83 & &\\
& 2010.71 & \ldots & \ldots & 100.6 & 0.08 & 1.00 & \ldots & 10.90 & &\\
& 2010.84 & \ldots & \ldots & 116.9 & 0.08 & 1.00 & \ldots & 10.96 & &\\\hline
\multirow{13}*{C\,3} & 1999.75 & 0.543 & 126.1 & \phantom{0}\phantom{0}3.2 & 0.20 & 1.00 & \ldots & \phantom{0}8.62 & \multirow{13}*{$\phantom{0}$$-5 \pm \phantom{0}7$} & \multirow{13}*{$-0.01 \pm 0.02$}\\
& 1999.85 & 0.688 & 129.3 & \phantom{0}\phantom{0}3.1 & \ldots & \ldots & \ldots & \phantom{0}\ldots & &\\
& 2000.02 & 0.774 & 133.6 & \phantom{0}\phantom{0}1.8 & 0.13 & 1.00 & \ldots & \phantom{0}8.77 & &\\
& 2000.22 & 0.491 & 134.6 & \phantom{0}\phantom{0}5.2 & 0.31 & 1.00 & \ldots & \phantom{0}8.44 & &\\
& 2008.41 & 0.401 & 141.8 & \phantom{0}10.1 & 0.30 & 1.00 & \ldots & \phantom{0}8.78 & &\\
& 2008.76 & 0.630 & 136.6 & \phantom{0}\phantom{0}7.7 & 0.22 & 1.00 & \ldots & \phantom{0}8.91 & &\\
& 2008.81 & 0.516 & 138.2 & \phantom{0}10.8 & 0.19 & 1.00 & \ldots & \phantom{0}9.21 & &\\
& 2009.15 & 0.728 & 138.3 & \phantom{0}\phantom{0}7.4 & 0.20 & 1.00 & \ldots & \phantom{0}8.97 & &\\
& 2009.42 & 0.446 & 135.5 & \phantom{0}11.6 & 0.15 & 1.00 & \ldots & \phantom{0}9.41 & &\\
& 2009.51 & 0.591 & 140.4 & \phantom{0}\phantom{0}8.3 & 0.14 & 1.00 & \ldots & \phantom{0}9.36 & &\\
& 2009.63 & 0.665 & 138.8 & \phantom{0}\phantom{0}9.4 & 0.32 & 1.00 & \ldots & \phantom{0}8.68 & &\\
& 2009.94 & 0.637 & 136.8 & \phantom{0}\phantom{0}8.0 & 0.20 & 1.00 & \ldots & \phantom{0}9.03 & &\\
& 2010.71 & 0.535 & 140.2 & \phantom{0}17.0 & 0.34 & 1.00 & \ldots & \phantom{0}8.89 & &\\
& 2010.84 & 0.554 & 138.4 & \phantom{0}10.7 & 0.16 & 1.00 & \ldots & \phantom{0}9.34 & &\\\hline
\multirow{13}*{C\,2} & 1999.75 & 1.257 & 135.4 & \phantom{0}\phantom{0}6.2 & 0.46 & 1.00 & \ldots & \phantom{0}8.19 & \multirow{13}*{$\phantom{0}\phantom{-}5 \pm \phantom{0}7$} & \multirow{13}*{$\phantom{-}0.01 \pm 0.02$}\\
& 1999.85 & 1.538 & 135.9 & \phantom{0}\phantom{0}3.6 & 0.13 & 1.00 & \ldots & \phantom{0}9.06 & &\\
& 2000.02 & 1.558 & 135.3 & \phantom{0}\phantom{0}3.2 & 0.29 & 1.00 & \ldots & \phantom{0}8.29 & &\\
& 2000.22 & 1.541 & 138.6 & \phantom{0}\phantom{0}5.0 & 0.44 & 1.00 & \ldots & \phantom{0}8.13 & &\\
& 2008.41 & 1.422 & 137.7 & \phantom{0}\phantom{0}6.5 & 0.55 & 1.00 & \ldots & \phantom{0}8.05 & &\\
& 2008.76 & 1.731 & 140.5 & \phantom{0}\phantom{0}4.9 & 0.43 & 1.00 & \ldots & \phantom{0}8.14 & &\\
& 2008.81 & 1.448 & 140.6 & \phantom{0}\phantom{0}6.4 & 0.43 & 1.00 & \ldots & \phantom{0}8.26 & &\\
& 2009.15 & 1.521 & 141.1 & \phantom{0}\phantom{0}5.0 & 0.40 & 1.00 & \ldots & \phantom{0}8.20 & &\\
& 2009.42 & 1.416 & 138.8 & \phantom{0}\phantom{0}6.9 & 0.56 & 1.00 & \ldots & \phantom{0}8.06 & &\\
& 2009.51 & 1.394 & 139.5 & \phantom{0}\phantom{0}7.2 & 0.55 & 1.00 & \ldots & \phantom{0}8.10 & &\\
& 2009.63 & 1.590 & 141.4 & \phantom{0}\phantom{0}6.7 & 0.73 & 1.00 & \ldots & \phantom{0}7.81 & &\\
& 2009.94 & 1.491 & 141.3 & \phantom{0}\phantom{0}6.9 & 0.50 & 1.00 & \ldots & \phantom{0}8.16 & &\\
& 2010.71 & 1.657 & 142.3 & \phantom{0}\phantom{0}7.7 & 0.66 & 1.00 & \ldots & \phantom{0}7.96 & &\\
& 2010.84 & 1.481 & 141.5 & \phantom{0}12.9 & 0.75 & 1.00 & \ldots & \phantom{0}8.07 & &\\\hline
\multirow{8}*{C\,1} & 1999.75 & 2.644 & 146.1 & \phantom{0}\phantom{0}4.5 & 0.76 & 1.00 & \ldots & \phantom{0}7.61 & \multirow{8}*{$\phantom{-}21 \pm 24$} & \multirow{8}*{$\phantom{-}0.06 \pm 0.07$}\\
& 1999.85 & 2.832 & 141.9 & \phantom{0}\phantom{0}3.1 & 0.73 & 1.00 & \ldots & \phantom{0}7.48 & &\\
& 2008.76 & 3.410 & 142.9 & \phantom{0}\phantom{0}3.7 & 0.75 & 1.00 & \ldots & \phantom{0}7.54 & &\\
& 2008.81 & 2.725 & 140.2 & \phantom{0}\phantom{0}3.1 & 1.18 & 1.00 & \ldots & \phantom{0}7.07 & &\\
& 2009.15 & 2.986 & 144.6 & \phantom{0}\phantom{0}5.5 & 1.60 & 1.00 & \ldots & \phantom{0}7.05 & &\\
& 2009.42 & 3.007 & 140.9 & \phantom{0}\phantom{0}3.2 & 1.10 & 1.00 & \ldots & \phantom{0}7.15 & &\\
& 2009.51 & 2.553 & 141.7 & \phantom{0}\phantom{0}2.7 & 0.34 & 1.00 & \ldots & \phantom{0}8.08 & &\\
& 2009.63* & 4.495 & 146.8 & \phantom{0}\phantom{0}4.5 & 1.80 & 1.00 & \ldots & \phantom{0}6.85 & &\\
& 2009.94 & 2.998 & 142.3 & \phantom{0}\phantom{0}3.3 & 1.44 & 1.00 & \ldots & \phantom{0}6.92 & &\\\hline
\end{tabular}
\tablefoot{An asterisk (*) indicates a component not used in the fit.\\
\tablefoottext{a}{Distance from core.}
\tablefoottext{b}{Position angle with respect to the core.}
\tablefoottext{c}{Flux density.}
\tablefoottext{d}{Major axis of fitted component.}
\tablefoottext{e}{Axial ratio of fitted component.}
\tablefoottext{f}{Position angle of component's major axis.}
\tablefoottext{g}{Log brightness temperature.}
\tablefoottext{h}{Apparent jet speed.}
\tablefoottext{i}{$\beta_\mathrm{app} = v_\mathrm{app}/c$}
}
\end{table*}

\begin{figure*}
\centering
\resizebox{\hsize}{!}{\includegraphics{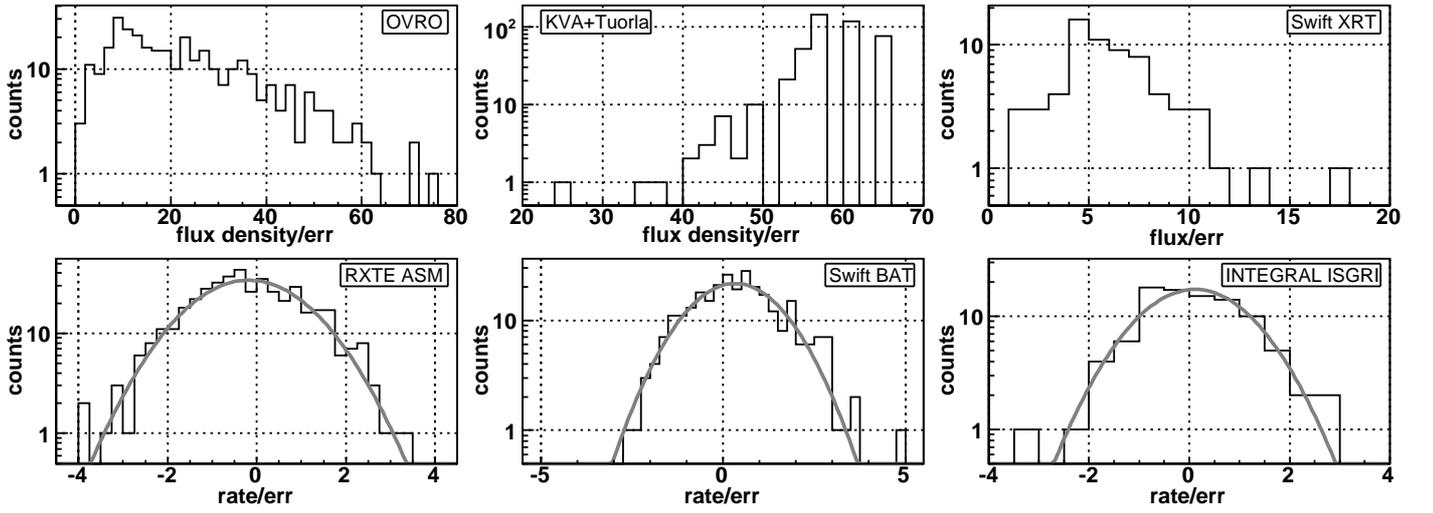}}
\caption{Distribution of flux density, flux or rate divided by the error for the individual light curves shown in Fig.~\ref{fig:2344_longterm}, for the last three panels including a fit with a Gaussian (shown in grey). See text for details.}\label{fig:2344_longterm_hist}
\end{figure*}

\begin{figure*}
\centering
\resizebox{\hsize}{!}{\includegraphics{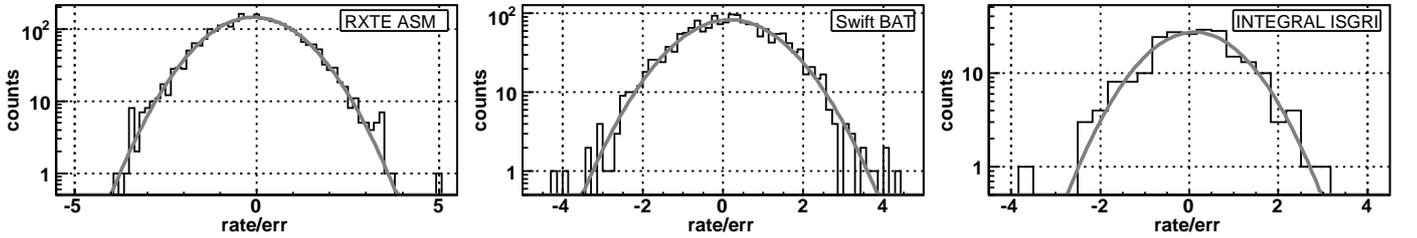}}
\caption{Distribution of rate divided by the error for the individual daily-binned light curves of \emph{RXTE} ASM, \emph{Swift} BAT and \emph{INTEGRAL} ISGRI including a fit with a Gaussian (shown in grey). See text for details.}\label{fig:2344_longterm_hist_daily}
\end{figure*}

\end{document}